\documentclass[12pt,preprint]{aastex}
\usepackage{appendix}
\usepackage{lscape}

\slugcomment{Revised version, to be submitted to the ApJ}

\begin{document}{}

\def\t0{\theta_{\circ}}
\def\muo{\mu_{\circ}}
\def\sd{\partial}
\def\be{\begin{equation}}
\def\en{\end{equation}}
\def\bv{\bf v}
\def\bvo{\bf v_{\circ}}
\def\ro{r_{\circ}}
\def\rhoo{\rho_{\circ}}
\def\etal{et al.\ }
\def\msun{\,M_{\sun}}
\def\rsun{\,R_{\sun}}
\def\rstar{\,R_{\star}}
\def\lsun{L_{\sun}}
\def\msunyr{{\rm M}_{\sun} {\rm yr}^{-1}}
\def\kms{\rm \, km \, s^{-1}}
\def\mdot{\dot{\rm M}}
\def\mdotd{\dot{\rm M}_{\rm d}}
\def\curf{{\cal F}}
\def\ecs{erg cm^{-2} s^{-1}}
\def \haebe{HAeBe}
\def \mum {\,{\rm \mu m}}
\def \simali {{\sim\,}}
\def \K {\,{\rm K}}
\def \Angstrom     {\,{\rm \AA}}
\newcommand \g            {\,{\rm g}}
\newcommand \cm           {\,{\rm cm}}

\title{CO J=2--1 Emission from Evolved Stars in the Galactic Bulge}

\author{Benjamin A. Sargent\altaffilmark{1,2},
N.~A. Patel\altaffilmark{3},
M. Meixner\altaffilmark{2},
M. Otsuka\altaffilmark{4},
D. Riebel\altaffilmark{5,6},
S. Srinivasan\altaffilmark{4,7}
}

\altaffiltext{1}{Space Telescope Science Institute, 3700 San Martin Drive, Baltimore, MD 21218, USA; {\sf baspci@rit.edu}}
\altaffiltext{2}{Center for Imaging Science and Laboratory for Multiwavelength Astrophysics, Rochester Institute of Technology, 54 Lomb Memorial Drive, Rochester, NY 14623, USA}
\altaffiltext{3}{Harvard-Smithsonian Center for Astrophysics, Cambridge, MA 02138, USA}
\altaffiltext{4}{Institute of Astronomy and Astrophysics, Academia Sinica, P.O. Box 23-141, Taipei 10617, Taiwan, R. O. C}
\altaffiltext{5}{Department of Physics and Astronomy, The Johns Hopkins University, 3400 North Charles St., Baltimore, MD 21218, USA}
\altaffiltext{6}{Department of Physics, United States Naval Academy, 572C Holloway Road, Annapolis, MD 21402, USA}
\altaffiltext{7}{UPMC-CNRS UMR7095, Institut d'Astrophysique de Paris, F-75014 Paris, France}

\begin{abstract}

We observe a sample of 8 evolved stars in the Galactic Bulge in the CO J=2--1 line using the Submillimeter Array (SMA) with angular resolution of 1--4 arcseconds.  These stars have been detected previously at infrared wavelengths, and several of them have OH maser emission.  We detect CO J=2--1 emission from three of the sources in the sample: OH 359.943 +0.260, [SLO2003] A12, and [SLO2003] A51.  We do not detect the remaining 5 stars in the sample because of heavy contamination from the galactic foreground CO emission.  Combining CO data with observations at infrared wavelengths constraining dust mass loss from these stars, we determine the gas-to-dust ratios of the Galactic Bulge stars for which CO emission is detected.  For OH 359.943 +0.260, we determine a gas mass-loss rate of 7.9 ($\pm$ 2.2) $\times$10$^{-5}\,\msunyr$ and a gas-to-dust ratio of 310 ($\pm$ 89).  For [SLO2003] A12, we find a gas mass-loss rate of 5.4 ($\pm$ 2.8) $\times$10$^{-5}\,\msunyr$ and a gas-to-dust ratio of 220 ($\pm$ 110).  For [SLO2003] A51, we find a gas mass-loss rate of 3.4 ($\pm$ 3.0) $\times$10$^{-5}\,\msunyr$ and a gas-to-dust ratio of 160 ($\pm$ 140), reflecting the low quality of our tentative detection of the CO J=2--1 emission from A51.  We find the CO J=2--1 detections of OH/IR stars in the Galactic Bulge require lower average CO J=2--1 backgrounds.

\end{abstract}

\keywords{circumstellar matter --- infrared: stars}

\section{Introduction}

At the end of a star's life, it returns material back to the interstellar medium from 
which it emerged.  This mass loss begins in the first red giant phase of a star's life 
\citep[e.g.,][]{groen12}.  Observations show gas mass loss from red giant stars 
\citep[][]{dup09}, but very little in the way of dust mass loss from red giants 
\citep[][]{mcd11}.  During the red giant phase, the mass loss is relatively weak, but 
it increases dramatically in the Asymptotic Giant Branch (AGB) phase.  The dust grains 
that form around AGB stars experience radiation pressure from the star, pushing them 
outwards, and they, in turn, push on the circumstellar gas, driving the gas outwards 
along with the dust \citep[e.g.,][]{for74}.  Detailed axisymmetric models by \citet{woitke06} 
confirm this picture for Carbon-rich (C-rich) AGB stars.  \citet{blahoe12} confirm this idea 
also works for Oxygen-rich (O-rich) AGB stars, which produce silicate dust, but require the 
dust grains to be Mg-rich, being highly transparent in the infrared, and to grow to 0.1--1 
$\mu$m sizes in order to have the large scattering cross-sections necessary to experience 
radiation pressure sufficient to push the grains outward to drive the mass outflow.  The 
mass-loss rate is an important parameter in characterizing the 
life of a star and also in constructing galaxy evolution models. For AGB stars, we 
have been estimating dust mass-loss rates through fitting optical through mid-infrared 
spectral energy distributions (SEDs; plots of flux versus wavelength) for stars in 
the Large Magellanic Cloud \citep[][]{riebel12} using radiative transfer models 
of emission from dust shells around stars \citep[e.g., see the ``GRAMS'' models 
constructed by][]{sarg11,srin11}.  However, dust mass loss is only a tiny fraction of 
the total mass loss for an AGB star.  \citet{knapp85} find an average gas-to-dust mass ratio 
for a sample of 22 OH/IR stars (stars detected in an OH maser and at infrared wavelengths) 
and Miras within 2-3 kpc of the Sun of $\simali$160.  Thus, in trying to measure an AGB star's 
{\it total} mass loss, it is better to measure the gas mass loss from the star.  
To translate dust mass-loss rates to gas mass 
loss rates, one must assume a gas-to-dust mass ratio; however, in 
many cases, the gas-to-dust mass ratio is unknown.  Therefore, gas mass-loss rates derived
from dust mass-loss rates often have large uncertainty.  Gas mass-loss rate measurements of 
additional AGB stars, in differing environments, 
are essential to improve the current situation, so that gas-to-dust ratios may be 
combined with dust mass-loss rates determined from SED-fitting (which can be 
done easily for large numbers of stars) to determine the total mass-loss rates for 
large populations of stars.

CO is very useful in determining gas mass-loss rates from evolved stars \citep[see][]{ram08,debeck10}.  \citet{winn09} observed a small sample of OH/IR stars close to the Galactic Center in the CO J=1--0 and J=2--1 lines.  Using Equation A.1 from \citet{ram08}, \citet{winn09} calculated the gas mass-loss rates for a couple of the stars in their sample.  CO J=2--1 and higher level transitions of CO, such as J=3--2, are useful in constraining the mass-loss rate history of AGB and OH/IR stars \citep[e.g.,][]{decin06}.  Gas mass-loss rates determined from CO line measurements \citep[][]{ram08} coupled with dust mass-loss rates determined for the same stars by modeling the dust emission seen in optical and infrared SEDs allows one to determine gas-to-dust ratios for such stars.  One may then compare these gas-to-dust ratios with metallicities, which may be determined from near-infrared (NIR) spectra of AGB stars, as was done by \citet{schul03}.

We follow \citet{winn09} and observe a sample of OH/IR and AGB stars in the inner Galactic Bulge \citep[also known as the Nuclear Bulge; see][]{mez96,schul03} in the CO J=2--1 line to investigate their gas mass-loss rates using the Submillimeter Array \citep[SMA;][]{ho04}.  The objects in the inner Galactic Bulge are in an extreme environment for star formation and evolution \citep[e.g., see][]{mez96}.  In particular, the typical metallicity of an inner bulge star is slightly higher than that of the local solar neighborhood, though inner bulge stars' metallicities range widely by a factor of at least 100 \citep[][]{schul03}.  This makes it an interesting place to study the variation of mass-loss rates with metallicity.  In addition, studies of inner Galactic Bulge AGB stars help remove the major uncertainty in mass-loss rate estimates of nearby AGB stars: the distance.  It should be noted that the inner Galactic Bulge is not a 2-dimensional structure in which all associated stars are exactly equidistant from Earth; instead, the inner Galactic Bulge is a 3-dimensional structure with characteristic radius of about 0.3 kpc \citep[][]{mez96}.  For simplicity, however, we adopt a distance to the inner Galactic Bulge of 8 kpc \citep[][]{reid93}.  There is the additional issue that, though our sample may be located near the Galactic Bulge on the sky, individual stars in the sample may not all be located 8.0 $\pm$ 0.3 kpc away.  This type of issue is specifically discussed by \citet{winn09} for both OH 359.762 +0.120 (also known as [SLO2003] E51) and OH 359.971 -0.119 and by \citet{blomm98} for OH 359.762 +0.120.

The line of sight to the inner Galactic Bulge contains intervening giant molecular clouds that confuse single dish detections of the AGB stars \citep[e.g., see][]{winn09,fong02}.  Also CO emission from background clouds may also confuse detections of these stars.  Interferometers are particularly good at filtering out extended structures, such as line of sight molecular clouds, in order to detect compact sources.

In this paper, we report on our SMA observations of OH/IR and AGB stars in the inner Galactic Bulge using the SMA.  In Section 2, we discuss our observations, and in Section 3, we discuss data reduction issues.  In Section 4, we present an analysis of our detections and non-detections.  In Section 5, we offer a discussion of gas-to-dust ratios and metallicities.  Finally, in Section 6 we draw conclusions from our study and present an outlook for future studies of mass loss from evolved stars.

\section{Observations}

We obtained our observations (see Table 1) using the SMA on Mauna Kea, Hawaii on May 3, 2010, May 11, 2010, September 25, 2010, and May 15, 2012 at a frequency of 230.538 GHz.  We also obtained observations on September 19, 2010 of the same targets observed September 25, 2010; however, we disregarded the September 19 observations as they were judged to be of lesser quality than the ones obtained for the same targets on September 25.  The array of 8 antennae was in the compact configuration for the May 2010 observations.  For the September 2010 and May 2012 observations, all antennae except antenna 6 were used, and these 7 antennae were in the extended configuration for the September 2010 observations and in the compact configuration for the May 2012 observations.  We give the exact FWHM beam major and minor axis values for each reported observation in its Figure caption.  Typical values are about 4.6$\arcsec$ and 2.7$\arcsec$, respectively, for the compact configuration observations and about 1.3$\arcsec$ and 1.1$\arcsec$, respectively, for the extended configuration observations.  At the frequency we observed our targets, 230.538 GHz, the field-of-view of our observations is 52$\arcsec$ across (the primary beam size).  We used the quasars 3c273, 3c279, and 3c454.3 as passband calibrators.  For gain/flux calibration, we used the quasars 1700-261, 1733-130, 1744-312, and 1802-396.

Our 2010 observations were conducted with the SMA correlator operating in the 1 receiver, 4 GHz mode, such that each channel had a spectral resolution of 812 kHz (corresponding to a velocity resolution of $\simali$1 km~s$^{-1}$ at 230.538 GHz frequency).  In this mode, the 4 GHz bandwidth is available simultaneously in both upper and lower sidebands; however, our analysis concentrates only on the upper sideband, as this is where our observations were tuned to place the 230.538 GHz line of CO J=2--1.  For our May 2012 observations, we used 1 receiver, 2 GHz mode, such that each channel had a spectral resolution of 406 kHz (velocity resolution of 0.5 km~s$^{-1}$).  The 2 GHz bandwidth is available simultaneously in both upper and lower sidebands, but, again, we concentrate only on the upper sideband to analyze the 230.538 GHz CO J=2--1 line.  The system temperature, T$_{\rm sys}$, would vary over each night of observing.  The ranges of T$_{\rm sys}$ were 80--300\,K for May 3, 2010; 80--210\,K for May 11, 2010; 100--200\,K for September 25, 2010 (outliers up to 1150\,K were eliminated in the data reduction process); and 100--300\,K for May 15, 2012. In the Figure caption of each contour plot, we give typical values of the RMS noise in the channel maps.

\section{Data Reduction}

The initial data reduction and calibration was done using the software MIR\footnote{https://www.cfa.harvard.edu/{\tt \~{}}cqi/mircook.html}.  After calibrating the data in MIR, the data were exported to the software Miriad\footnote{http://www.cfa.harvard.edu/sma/miriad/} \citep[][]{sault95} for output of spectra and synthesis of images.  In Miriad, we used the {\it invert} task to transform visibilities to channel maps.  Most maps are 67$\times$67 pixels in size, with the square pixels being $\simali$0.6$\arcsec$ on a side.  Next, we applied a hybrid Hogbom/Clark/Steer {\it clean} algorithm.  This was followed by applying the {\it restor} task, which generated a cleaned map for each channel from the {\it clean} task output.  We then used the {\it cgdisp} task to overlay the restored channel contour maps on top of {\it Spitzer} GLIMPSE \citep[][]{ben03,chur09} images of the same regions.  The output cleaned maps were searched for point sources.  The spectrum of each point source was obtained by using the {\it imspec} task to compute the spectrum for an aperture (a box usually about 3$\arcsec$ $\times$ 4.8$\arcsec$ in size) at the position of the point source.  When a point source was found, the channels where the point source emission was strongest were binned together, and the centroid of the point source in the binned image was determined.  We report the coordinates of the centroid position of each of the detected sources in our sample in Table 2.

Guided by plots of visibility amplitude versus projected baseline length, \citet{winn09} remove the shorter baselines from their SMA data in order to remove interstellar emission contamination from their data.  Our data for OH359 was already fairly free of interstellar contamination, so we decided not to remove baselines for this source's data.  For A12 and A51, point sources were visible without removing baselines but were heavily contaminated by extended interstellar emission.  Accordingly, we remove baselines $<$ 25 k$\lambda$ from our A12 data, and we remove baselines $<$ 20 k$\lambda$ from our A51 data.  The improvement in the images and spectra for these two sources was dramatic.  We investigated removing baselines for the rest of our sample, but doing so did not improve any of their images or spectra, so we opted not to remove any of their baselines for the rest of the data we present here.

The distance between the center of the SMA field-of-view and the centroid position was then used to determine the correction scalar for the output spectrum.  The correction scalar is determined by assuming a gaussian, centered at the center of the field-of-view, to have a full-width at half maximum (FWHM) of 52$\arcsec$.  The size of the primary beam is indicated in the contour plots as the magenta circle (or arcs, if the plot is zoomed in).  The correction scalar is the inverse of the value of the gaussian at the position of the centroid.  This is the primary beam correction.

In order to know at which velocity to expect the CO J=2-1 emission, the radial velocity for a source is needed.  We gathered the radial velocities of six of the OH/IR stars in our sample from the OH 1612 MHz maser study by \citet{sjou98}.  The radial velocity for the other OH/IR star, A29, was obtained from the OH 1612 MHz maser study by \citet{lqv92}.  No radial velocity was available for the AGB star A27.

We determined the positional accuracy for our SMA data by using the Miriad {\it imfit} task to fit a 2-dimensional gaussian to images of the quasars we observed as passband calibrators (3c279, 3c273, and 3c454.3) and computing the positional offset between the best-fit 2-d gaussian and the center of the SMA field-of-view.  The average value for this offset was about 0.7$\arcsec$.  This can be compared to the positional accuracy determined for point sources detected by the GLIMPSE team\footnote{http://www.astro.wisc.edu/sirtf/glimpse360\_dataprod\_v1.1.pdf} using {\it Spitzer} of $\simali$0.3$\arcsec$.  Thus, the relative positional accuracy between our SMA data and {\it Spitzer} GLIMPSE images would be the sum in quadrature of these two accuracies, or about 0.8$\arcsec$.

\section{Analysis}

We detected a point source in the channel maps at the expected position of OH 359 (Figure 1), and a point source is detected at 8$\mum$ in a {\it Spitzer} GLIMPSE image.  The channel maps spanning the range of OH maser emission for this object \citep[for this target's OH maser velocity range, see Table 2 of][]{sjou98} were binned, and the centroid of the point source (green cross in Figure 1) as seen in this velocity-binned image was set as the center of a 3.44$\arcsec \times$4.8$\arcsec$ box in which the spectrum was extracted using the MIRIAD routine {\it imspec}.  The spectrum of OH359 showed, as did the spectra from all the SMA observations of our sample, CO emission in the velocity range between about -180 km~s$^{-1}$ and +180 km~s$^{-1}$ (see also the discussion in Section 5.2).  This box size was set a little larger than the beam size for this observation, in order to optimize the performance of {\it imspec}.  We applied the primary beam correction (described above) to this spectrum, as did all CO spectra shown in this paper.  We followed \citet{winn09} and fit a parabola,

\begin{eqnarray}
F_{\nu} & = & 
       F_{\nu, \rm max}[1-(\frac{V-V_{rad}}{V_{exp}})^{2}]  ~~,
\end{eqnarray}

\noindent to the CO J=2--1 line in the resulting spectrum of OH359, such that F$_{\nu, \rm max}$ is the flux density at the peak of the parabola, V$_{rad}$ is the CO J=2-1 line center velocity, V$_{exp}$ is half the width of the parabola at the baseline.  This model assumes unresolved, optically thick emission from the CO gas giving rise to the J=2--1 line.  The center, width, and peak flux of the parabola are allowed to be free parameters.  The best fit, shown in Figure 2, has the parameters F$_{\nu, \rm max}$ = 0.36 Jy, V$_{rad}$ = -182.3 km~s$^{-1}$, and V$_{\rm exp}$ = 19.1 km~s$^{-1}$.  Note the good matching between the velocity range of the best-fit model and the red and blue limits of the OH maser emission detected by \citet{sjou98}.  We then use the formula presented by \citet{ram08} to obtain total mass-loss rate from the integrated line flux of the CO J=2-1 line.  To obtain integrated flux in units of K km~s$^{-1}$ as required for the \citet{ram08} formula, we translated our spectra from flux density to brightness temperature using the Planck formula.   We find a total mass-loss rate of 7.9 ($\pm$ 2.2) $\times$10$^{-5}\,\msunyr$ for OH359, assuming a CO-to-H$_{2}$ abundance ratio of 2$\times$10$^{-4}$ \citep{winn09} and an adopted distance of $\simali$8 kpc \citep[][]{reid93}.  A caveat, as noted by \citet{winn09} and \citet{ram08}, is that the \citet{ram08} mass-loss rate formula has only been checked to be valid for mass-loss rates between 10$^{-7}$ and 10$^{-5}\,\msunyr$.  Also, the \citet{ram08} mass-loss rate formula applies only for unresolved emission, so a further caveat is that if any signal from our sources falls outside the rectangular regions from which we extracted spectra (whether due to sidelobes or due to the source intrinsically being extended/resolved), then the extracted fluxes will be artificially lowered, and the calculated gas mass-loss rate will be correspondingly lowered.

In assigning errors to the parabola fit parameters, instead of using the standard deviation on a quantity, we initially used the Median Absolute Deviation (MAD), as \citet{riebel12} used when determining uncertainties on dust mass loss rates.  For the parameter in question, this entails taking the median value of the parameter in the N best-fit models (i.e., the models with the N lowest chi-squared values), finding the absolute value of the difference between each model's parameter value and the median parameter value, and then finding the median of the set of absolute values of differences.  As \citet{riebel12} note, for normally distributed errors, this MAD uncertainty is somewhat lower than the standard deviation.  Instead of determining MAD from the 100 best models, as \citet{riebel12} did, we take the 10\,000 best models, as we explore a grid of $>$ 10$^{6}$ models (all combinations of 101 values of each of F$_{\nu, \rm max}$, V$_{rad}$, and V$_{\rm exp}$) to model a given CO line, in comparison to the $\simali$12\,000 (Carbon-rich) or $\simali$69\,000 (Oxygen-rich) GRAMS models \citet{riebel12} explored when modeling evolved star SEDs.  We determined the MAD uncertainties on F$_{\nu, \rm max}$, V$_{rad}$, and V$_{\rm exp}$; however, these uncertainties seemed low.  The OH359 spectrum (Figure 2) shows single-channel (1 km~s$^{-1}$) scatter, so we assign 1 km~s$^{-1}$ uncertainty to the radial velocity, V$_{rad}$.  Since the expansion velocity involves measuring the location of both the red and blue ends of the CO line, and each such measurement has the same uncertainty, we assign to the expansion velocity, V$_{\rm exp}$ (which is a difference in velocities; i.e., the red and blue ends of the CO line), the square root of 2 times the uncertainty assigned to the radial velocity, V$_{rad}$.  Thus, we assign an uncertainty on the expansion velocity for OH359 of 1.4 km~s$^{-1}$.  We use the MAD uncertainty for the line strength, F$_{\nu, \rm max}$.  We propagate these errors through our calculations of the integrated CO line intensity, I$_{\rm co}$, and the gas mass-loss rate, $\mdot$ (see Table 2 for all these uncertainties).

We also detected a point source at the expected position of A12 (Figure 3), and a point source is detected in an 8$\mum$ {\it Spitzer} GLIMPSE image.  We bin the channel maps between velocities of 95 and 117 km~s$^{-1}$ to determine the centroid of the CO emission.  We obtain the spectrum with {\it imspec} using a 3.45$\arcsec \times$2.01$\arcsec$ box centered at the centroid of the point source in the velocity-binned image.  We obtain the parameters F$_{\nu, \rm max}$ = 0.91 Jy, V$_{rad}$ = 105.0 km~s$^{-1}$, and V$_{\rm exp}$ = 14.3 km~s$^{-1}$ from the best-fitting model (Figure 4).  Here, the matching between the velocity range of our CO model and the OH maser detection by \citet{sjou98} is not as good as for OH359.  The OH emission is likely less contaminated by interstellar emission than the CO is, so we focus on how the intrinsic CO profile of this star might be affected by interstellar CO emission.  Removing short baselines should preferentially eliminate extended emission structures in the data but may also eliminate emission arising from compact sources.  It is possible there is still some extended emission that has not been eliminated by baseline removal, and it is also possible that emission from the point source has been over-subtracted.  For these reasons, the CO emission seen in Figure 4 may not be exactly the total circumstellar CO emission from A12; in other words, the CO emission seen in Figure 4 may not be identical to how it would appear if A12 did not have the contaminating interstellar CO emission.  For A12, because the peaks and dips in the noise both outside of the CO line and on top of the CO line are all about 3 channels wide, we assign uncertainties of 3 km~s$^{-1}$ to V$_{rad}$ and the square root of 2 times this, or 4.2 km~s$^{-1}$, to V$_{\rm exp}$, while retaining the MAD uncertainty on F$_{\nu, \rm max}$.  Again we use the \citet{ram08} formula to obtain total mass-loss rate from the integrated line flux of the CO J=2--1 line, and we find a gas mass-loss rate of 5.4 ($\pm$ 2.8) $\times$10$^{-5}\,\msunyr$ for A12, with the same caveats as for OH359.

Additionally, we detected a point source at the expected position of A51 (Figure 5), and a point source is found in an 8$\mu$m {\it Spitzer} GLIMPSE image centered on this star.  Here, we bin the channel maps between -101 and -87 km~s$^{-1}$ to determine the centroid of the CO emission point source.  We obtain the spectrum using a 3.65$\arcsec \times$1.87$\arcsec$ box centered at the centroid of the point source in the velocity-binned image.  The model that best fits this data (Figure 6) has the parameters F$_{\nu, \rm max}$ = 3.1 Jy, V$_{rad}$ = -94.8 km~s$^{-1}$, and V$_{\rm exp}$ = 6.0 km~s$^{-1}$.  For A51, there is a marked discrepancy between the velocity ranges of the CO emission model and the OH maser detection by \citet{sjou98}.  Again, it is possible the baseline removal affected the shape of the CO line.  Though we find V$_{\rm exp}$ = 6.0 km~s$^{-1}$ for A51, lower values of V$_{\rm exp}$ for O-rich AGB stars have been reported in the literature \citep[see the terminal velocity of 4.2 km~s$^{-1}$ for V438 Oph given by][]{debeck10}.  We assign the same uncertainties to the radial and expansion velocities of A51 as we did those of A12 for the same reasons and again retain the MAD uncertainty on F$_{\nu, \rm max}$.  The \citet{ram08} formula gives a gas mass-loss rate for A51 of 3.4 ($\pm$ 3.0) $\times$10$^{-5}\,\msunyr$, again with the same caveats as for OH359.

However, because of the large discrepancy between the V$_{\rm exp}$ we determine from CO emission and the velocity range between the red and blue peaks of the OH maser emission for this source \citep[][]{sjou98}, we consider this at best only a tentative detection of CO J=2--1 emission from A51.  \citet{olof00} observed that the carbon star TT Cyg has a detached shell.  The spectrum they obtained for this source shows a relatively narrow peak indicating an expansion velocity of only a few km~s$^{-1}$, similar to A51; unlike A51, however, TT Cyg's spectrum shows evidence for a detached shell in addition to the central peak.  \citet{olof00} conclude this is due to a recent mass-loss episode, and the detached shell is due to an older episode.  It is possible A51 has a similar mass-loss history, though we do not find a second CO emission peak indicating a detached shell for A51.  \citet{hab96} note that CO emission from the circumstellar envelope around an AGB star should arise further away from the star than OH emission.  From this separation of the regions where OH and CO emission is produced in AGB envelopes, it is conceivable the OH has a higher expansion velocity than the CO.  Again, it is also possible that the difference in expansion velocities determined from OH emission versus CO emission is not real, but only an artifact of heavy contamination by interstellar emission that has either been incorrectly removed in data reduction, whether by over- or under-subtraction.  A51 deserves further study, perhaps using the more sensitive ALMA array, to be certain whether its CO J=2--1 emission has been detected.

The parameters of the best-fit parabolic models of OH359, A12, and A51 are given in Table 2, along with the computed gas mass-loss rates.

\subsection{Non-Detections}

In addition to OH359, A12, and A51, we also observed a number of fields where we expected to find our other targets.  For none of the other targets were compact sources seen in the center of the fields-of-view of the SMA maps.  However, compact sources were seen away from the centers of the fields of view in some of our SMA data.  In the A10 data, we found a compact source to the northwest of the center of the field of view.  In searching the infrared image (Figure 7) at the position of the compact source, we found a faint point source in the GLIMPSE image, about 1$\arcsec$ from the centroid of the compact source seen in the SMA data.  The SMA spectrum of this source (Figure 8) shows strong emission at a line center near 96 km~s$^{-1}$.  A parabolic fit to this line finds F$_{\nu, \rm max}$ = 2.3 Jy, V$_{rad}$ = 96 km~s$^{-1}$, and V$_{\rm exp}$ = 11 km~s$^{-1}$.  After constructing an SED using Vizier in SIMBAD and fitting a GRAMS O-rich model \citep[][]{sarg11} to its infrared SED to obtain a dust mass-loss rate, and after using the \citet{ram08} mass-loss rate formula to obtain a total mass-loss rate from the parabolic model fit, the implied gas-to-dust ratio would be about 3$\times$10$^{6}$.  This, combined with its infrared SED appearing almost like that of a naked stellar photosphere suggests that the CO emission from the compact source centered at 96 km~s$^{-1}$ is not likely from an OH/IR or other dusty evolved star.  Note that the central velocity of this CO emission, 96 km~s$^{-1}$, is well within the range of interstellar emission seen in all our Galactic Bulge SMA spectra of -180 km~s$^{-1}$ and +180 km~s$^{-1}$ (see Section 4 and 5.2).  Also, because it is so far from the coordinates of A10, it is not likely a detection of CO emission from A10.  The off-center compact source in the A10 field may arise instead from a compact clump of CO in the interstellar medium along the line of sight to A10 \citep[again, see discussion by][]{winn09}.

A similar situation was found for a compact source seen off-center in the A12 field (without baselines removed).  The SED for the faint source seen in the infrared image about 1 $\arcsec$ away from the compact source seen in the SMA data (Figure 9) appears photospheric in shape.  At 181 km~s$^{-1}$ in the A12 data, one finds a strong peak of emission arising in the SMA data (Figure 10) for an off-center compact source.  A parabolic fit to this line finds F$_{\nu, \rm max}$ = 3.2 Jy, V$_{rad}$ = 181 km~s$^{-1}$km~s$^{-1}$, and V$_{\rm exp}$ = 9 km~s$^{-1}$.  Note that the central velocity of this CO emission, 181 km~s$^{-1}$, is well within the range of interstellar emission seen in all our Galactic Bulge SMA spectra of -180 km~s$^{-1}$ and +180 km~s$^{-1}$ (see Section 4 and 5.2).  In addition, the gas-to-dust ratio that would be implied for this object by GRAMS O-rich modeling of the SED and by using the \citet{ram08} formula after fitting a parabolic model to the line seen in the SMA spectrum is also unrealistically high, about 10$^{5}$.  Therefore, the compact source seen around 180 km~s$^{-1}$ in the SMA data is not likely from an OH/IR or other dusty evolved star.

For A27, we were not aware of an OH maser detection to guide us in looking for its CO emission.  We did not find a compact source of CO emission at the coordinates for this star.  However, we did find a compact source to the southwest of its coordinates, centered at a radial velocity of about 113 km~s$^{-1}$ (Figure 11).  We extracted the spectrum at the position of the green cross in Figure 11, fitting a parabola to the line at 113 km~s$^{-1}$, and we find F$_{\nu, \rm max}$ = 2.6 Jy, V$_{rad}$ = 113 km~s$^{-1}$, and V$_{\rm exp}$ = 10 km~s$^{-1}$.  Unfortunately, we did not find a detection at this position of a point source in any infrared catalogs, so we find it unlikely the compact source in the A27 field is associated with an evolved star.

For the other three targets in our sample, A29, A52, and A7, we detected neither them nor any off-center compact sources.

\section{Discussion}

\subsection{Gas/Dust Ratios}

\citet{knapp85} find an average gas-to-dust mass ratio for nearby Oxygen-rich (O-rich) AGB stars of 160.  The {\it dust-to-gas} mass ratio for AGB stars may vary linearly with metallicity \citep[][]{marsh04,vl06}, though this is not totally clear \citep[][]{vl06}.  \citet{schul03} find that the Galactic Bulge should, on average, have a slightly lower metallicity than the Sun ([Fe/H] = 0 dex) but a slightly higher metallicity than the solar neighborhood \citep[see Figure 9 of][]{schul03}.  This suggests that the gas-to-dust mass ratio of the Galactic Bulge overall should be $\leq$ 160.  However, the metallicities for the Galactic Bulge stars range widely, extending to very low metallicities \citep[][]{schul03}, so the lowest metallicity Galactic Bulge stars may be expected to have gas-to-dust ratios $>$ 160.  Our four detected stars all have subsolar metallicities (Table 3).

The infrared SEDs of OH359, E51, A12, and A51 were assembled by collecting their photometry using the Vizier tool in SIMBAD\footnote{see http://simbad.u-strasbg.fr/simbad/}.  Table 3 lists the relevant information for how we collected this photometry.  The {\it Spitzer} IRAC and MIPS 24 $\mu$m photometry were then de-reddened using the relative extinction points plotted as diamonds and labeled as ``\citet{flah07}'' in the bottom panel of Figure 1 of \citet{mcc09}.  All other photometry were dereddened by interpolation (evaluated at the isophotal wavelengths of the bands) of the \citet{mcc09} 1 $\leq$ A$_{\rm K}$ $<$ 7 relative extinction curve.  All relative extinctions are normalized using A$_{\rm K}$ determined from the star's A$_{\rm V}$ value (see Table 3) and the \citet{mcc09} 1 $\leq$ A$_{\rm K}$ $<$ 7 relative extinction curve \citep[which, for optical and near-infrared wavelengths, is the R$_{\rm V}$=5.0 curve of][]{math90}.  For E51, A12, and A51, we use the A$_{\rm V}$ value from \citet{schul03}; for OH359, we use E($H$-$K$) = 0.97 for target \#64-28 of \citet{wood98} to determine A$_{\rm K}$, which we then use to determine A$_{\rm V}$, assuming the R$_{\rm V}$ = 5.0 curve of \citep[][]{math90}.  The dereddened SEDs for OH359, E51, A12, and A51 are given in Figures 13, 14, 15, and 16, respectively.

We did not fit Carbon-rich (C-rich) GRAMS models \citep[][]{srin11} to these 4 Galactic Bulge star SEDs because OH masers have been detected for each of these 4 stars \citep[for OH359, A12, and A51, see Table 1; for E51, see][]{winn09}, so we conclude they are likely O-rich.  We therefore fit O-rich GRAMS models to the dereddened SEDs of these stars, after the manner of \citet{riebel12}.  Using the measurements of these 4 stars' variability (see Table 3), we determined a value of $\sigma_{var}$ to use in inflating the error bars of all photometry \citep[again, after the manner of][]{riebel12} except for the {\it Spitzer}-IRAC photometry from GLIMPSE, which we treat as a reference set of photometry.  We treat the \citet{rmz08}, GLIMPSE, and WISE photometry as belonging to 3 different epochs, fitting each epoch's SED separately.  Any $JHK_{s}$, AKARI, or MIPS-24 photometry for a star is treated as common to the SEDs of all 3 ``epochs'' for that star.  For OH 359, the 3.6 and 4.5 $\mu$m photometry from \citet{rmz08} and GLIMPSE agree fairly well with the WISE 3.4 and 4.6 $\mu$m photometry; however, the 5.8 and 8.0 $\mu$m \citet{rmz08} and GLIMPSE fluxes are about 3 times higher than it appears they should be, compared to the rest of the \citet{rmz08}, GLIMPSE, and WISE photometry, so they were omitted from the SED-fitting.  The discrepant photometry at 5.8 and 8.0 $\mu$m may be due to the ``bandwidth effect''\footnote{see http://irsa.ipac.caltech.edu/data/SPITZER/docs/irac/iracinstrumenthandbook/}, which results in artifacts appearing in 5.8 and 8.0 $\mu$m data as point sources centered multiples of 4 pixels away from the star \citep[see also][]{rmz08}, if it is a bright star (see Figure 1).  The image we show in the background of Figure 1 has been processed so that the pixels are 0.6$\arcsec$ long, as opposed to 1.2$\arcsec$ in the original data, so the artifact point sources show up in Figure 1 as multiples of 8 pixel away from the star.  If not corrected, this artifact might affect 5.8 and 8.0 $\mu$m fluxes for affected targets.

We obtain dust mass-loss rates of 2.5 ($\pm$ 0.2) $\times$10$^{-7}\,\msunyr$, 8.4 ($\pm$ 1.1) $\times$10$^{-7}\,\msunyr$, 2.4 ($\pm$ 0.1) $\times$10$^{-7}\,\msunyr$, and 2.1 ($\pm$ 0.2) $\times$10$^{-7}\,\msunyr$ for OH359, E51, A12, and A51, respectively, from the average of the best-fit model to each star's SED of the 3 ``epochs''.  For A51, GLIMPSE photometry was not available, so there were only 2 ``epochs'' for A51.  Note that this takes into account that these stars' dust envelope expansion velocities are not 10 km~s$^{-1}$, as assumed by \citet{sarg11}; rather, the OH359, A12, and A51 dust expansion velocities are assumed to be equal to the gas expansion velocities as observed by us (Table 2), and the E51 dust expansion velocity is assumed to be the same as that reported for CO by \citet{winn09}.  This assumption that the dust expansion velocity is equal to the gas expansion velocity for AGB star envelopes is only an approximation, as models that allow for drift of dust with respect to gas in these envelopes show that the dust may have slightly higher expansion velocities than the gas \citep[e.g.,][]{sanhoe03}.  However, as our dust models are always quite optically thick at 10 $\mu$m wavelength (note all the models seen in Figures 13-16 have absorption in the 10 $\mu$m feature), which means they are also quite optically thick at optical wavelengths and have relatively large gas mass-loss rates (few times 10$^{-5}\,\msunyr$; Table 2), these stars' envelopes should not  be in what \citet{ivel10} called the ``drift-dominated regime''.  Thus, drift of dust through the gas is less significant for the stars we discuss than for AGB stars with lower mass-loss rates.  We estimate uncertainties on the dust mass-loss rates for these 4 stars (Table 3) in the same way \citet{riebel12} estimated uncertainties on their dust mass-loss rates; that is, using the MAD uncertainties determined from the 100 best-fit models (again, here drawing only from the Oxygen-rich GRAMS models).  This results in gas-to-dust ratios of about 310 ($\pm$ 89), 220 ($\pm$ 110), and 160 ($\pm$ 140) for OH359, A12, and A51, respectively.

The gas mass-loss rate for E51 is 6$\times$10$^{-5}\,\msunyr$ (A. Winnberg, private communication).  We estimate the uncertainty on the E51 gas mass-loss rate by propagating the CO J=2--1 line F$_{\nu, \rm max}$, V$_{rad}$, and V$_{\rm exp}$ 1-$\sigma$ uncertainties given by \citet{winn09} for E51.  From this, we estimate the gas mass-loss rate uncertainty for E51 to be about 2$\times$10$^{-5}\,\msunyr$.  We then determine a gas-to-dust ratio for E51 of 71 ($\pm$ 23).  The dust mass-loss rates and gas-to-dust ratios for all 4 stars for which we modeled the dust shells using GRAMS are also listed in Table 3.

As \citet{schul03} note, Solution 1 from \citet{ram00} is a formula to compute metallicity, [Fe/H], for stars using observations of various lines seen in near-infrared spectra.  Using this formula and the equivalent widths of various near-infrared lines reported by \citet{schul03}, A12 is found to have a metallicity, [Fe/H], of -0.82 dex, E51 is found to have [Fe/H] = -0.92, and A51 is found to have [Fe/H] = -0.38.  \citet{schul03} note the typical errors on metallicities determined by this method to be about 0.1 dex.  Using equivalent widths of the same near-infrared lines reported by \citet{vanhol06}, OH359 is found to have [Fe/H] = -0.81.  The distribution of metallicities of the Galactic Bulge star sample of \citet{schul03} spans [Fe/H] from 0.5 down to $<$ -2.0 dex, with the histogram of metallicities peaking around -0.3 dex.  \citet{schul03} note the average metallicity of the local solar neighborhood sample of \citet{lanwood00} is [Fe/H] $\sim$ -0.6.  Thus, the metallicities of OH359, E51, and A12 are quite lower than solar, and all of them are slightly lower than that of the solar neighborhood.  It is unknown why the gas-to-dust ratio for E51 is so low, especially considering the low metallicity of E51.  The average of the gas-to-dust ratios of all 4 stars for which CO J=2--1 is detected - OH359, A12, and A51 (by us) and E51 \citep[by][]{winn09} - is about 190.  This may be understood to be larger than the Galactic Bulge upper limit of $\leq$ 160 (see discussion at the beginning of this subsection) due to the low metallicities of these stars as determined from their near-infrared spectra.  It is also possible the dust mass-loss rates assumed for the gas-to-dust ratio calculations are uncertain due to such factors as these stars' variability, differing dust properties than those assumed for our modeling, etc.  We have obtained ground-based photometry and spectroscopy at near- and mid-infrared wavelengths of a number of our Galactic Bulge targets that we will use to constrain further the dust mass-loss rates and the dust properties of these stars (Sargent et al, {\it in prep}).

We do note that the gas-to-dust ratios of OH359, E51, A12, and A51 are consistent with the gas-to-dust ratios determined by \citet{groen99} for O-rich and C-rich AGB stars in the Milky Way within about 8 kpc of Earth.  According to \citet{jones94}, the period of E51 is 715 days, and according to \citet{wood98}, the periods of OH359, A12, and A51 are 692, 552, and 759 days, respectively.  The gas-to-dust ratios we determine for these 4 stars are all consistent with the data in the plot of the logarithm of gas-to-dust ratio versus logarithm of period in Figure 20 of \citet{groen99}.

However, we caution against concluding too much regarding these gas-to-dust ratios, considering they are for a very small sample of 4 stars.  We also note that the lower quality detections of A12 and (especially) A51 seem to result in relatively larger uncertainties on their gas-to-dust ratios.  The uncertainty on the A51 gas-to-dust ratio seems to be dominated by the relatively large uncertainty on its gas mass-loss rate.  This uncertainty probably arises largely from the huge uncertainty we assign to the expansion velocity, 6.0 $\pm$ 4.2 km~s$^{-1}$, which is due to the poorer data quality for this star's SMA spectrum.  For A51, we suggest follow-up observations of its CO emission would establish firmly its detection and better constrain its gas mass-loss rate.

As noted in the introduction, the goal of these studies of evolved star CO emission is to constrain their gas-to-dust ratios, as there are few determinations of gas-to-dust ratios of evolved stars in the literature.  The Galactic Bulge is an ideal place to make such determinations, as the distance to the stars is well-known ($\simali$8 kpc), which reduces uncertainty in these stars' other properties (e.g., luminosity).  By determining reliable gas-to-dust ratios for these stars, for which we also know metallicity \citep[see][]{schul03,vanhol06}, we hope to be able to determine gas-to-dust ratios for lower-metallicity environments, like the Magellanic Clouds.  We note specifically that OH359, E51, and A12 have metallicities very similar to that of the Small Magellanic Cloud of [Fe/H] = -0.64 \citep[][]{russ92,roll99}, and A51 has a metallicity quite similar to that of the Large Magellanic Cloud, which has [Fe/H] = -0.30 \citep[][]{russ92}; therefore, our work applies to studies of mass loss in the Magellanic Clouds.  This will aid in studies of the life cycle of matter of galaxies, by constraining the contribution of matter to the galaxy from its evolved star population.

\subsection{Lessons Learned for Future Searches}

We noticed the spectra of all the targets in our sample included a lot of extended emission roughly between -180 km~s$^{-1}$ and +180 km~s$^{-1}$, not associated with any of our evolved star sample.  However, outside of this velocity range, the emission was usually much weaker.  \citet{winn09} similarly noted that significant interstellar background emission is usually found between -200 km~s$^{-1}$ and +200 km~s$^{-1}$.  Pursuing such high radial velocity sources may prove a successful means of detecting CO emission at J=2--1 from evolved stars than the lower velocity sources, as our detection of OH359 demonstrates.  In the \citet{sjou98} sample alone, there are 7 OH maser sources with central OH maser velocities $<$ -180 km~s$^{-1}$ or $>$ +180 km~s$^{-1}$.  It should be noted, however, that \citet{winn91} took this sort of approach with the IRAM 30m telescope in the CO J=2--1 and CO J=1--0 lines for a sample of 17 stars with moderately high radial velocities (greater than +100 km~s$^{-1}$ or less than -100 km~s$^{-1}$) and only detected 2 of them.

Observing CO emission at higher J $\rightarrow$ J-1 transitions should also prove fruitful for detecting CO emission from Galactic Bulge stars.  We estimated the CO J=3--2, J=4--3, and J=6--5 peak line fluxes for OH359, A12, and A51 based on our CO data and for E51 based on the \citet{winn09} CO data.  Using the mass-loss rates we determine for OH359, A12, and A51, and the rate of 6$\times$10$^{-5}\,\msunyr$ for E51 (from A. Winnberg, private communication), we compute CO J=3--2, CO J=4--3, CO J=6--5 peak fluxes using ratios of main beam brightness temperature for O-rich AGB stars whose fitted line profile $\beta$ parameter is between 1 and 3 in the study by \citep[][]{debeck10}.  For OH359, the J=3--2, J=4--3, and J=6--5 estimated peak line fluxes are 1.8 Jy, 0.28 Jy, and 9.1$\times$10$^{-6}$ Jy, respectively; for E51, the flux estimates are 3.4 Jy, 0.93 Jy, and 4.3$\times$10$^{-4}$ Jy, respectively; for A12, the flux estimates are 1.7 Jy, 0.25 Jy, and 6.5$\times$10$^{-6}$ Jy, respectively; and for A51, the flux estimates are 9.1 Jy, 4.8 Jy, and 0.069 Jy, respectively.  The predicted CO J=3--2 and J=4--3 fluxes are larger than of comparable to the CO J=2--1 fluxes of 0.36 Jy for OH359, 0.6 Jy for E51 \citep[][]{winn09}, 0.91 Jy for A12, and 3.1 Jy for A51; however, the predicted CO J=6--5 fluxes are much lower.

\citet{saw01} find the CO J=2--1 to CO J=1--0 intensity ratios for the interstellar CO line emission in the central 900 pc of the Galaxy near unity.  \citet{oka12} find the CO J=3--2 to CO J=1--0 intensity ratio, $R_{3-2/1-0}$, in the Galactic Center for the interstellar CO line emission to be about 0.7.  Additionally, \citet{mart04} find CO J=4--3 to CO J=1--0 intensity ratios for the Galactic Center interstellar CO line emission of about unity, which they note is consistent with the findings of \citet{kim02}.  As \citet{just96} reported the observed and model antenna temperatures (and integrated line intensities) for OH 26.5+0.6 to rise dramatically as one progresses to higher CO J $\rightarrow$ J-1 transitions, the utility of observing Galactic Bulge stars in higher CO transitions becomes apparent.

We also note that the maps of Galactic Center CO J=1--0 emission by \citet{oka98} and CO J=2--1 emission by \citet{saw01} are consistent with the detections and non-detections of the stars in our target list and E51 \citep[by][]{winn09}.  OH359, which we cleanly detected, is located completely outside of CO emission in these maps.  E51 \citep[detected by][]{winn09} and A12 are at the edges of CO emission clouds seen in the \citet{oka98} and \citet{saw01} maps.  A51 is located in weak CO background emission.  The 5 sources that were not detected are located amongst significant CO emission in these maps.  OH359, E51, A12, and A51 are all located where the background CO J=2--1 mean intensity in the \citet{saw01} maps is $\leq$ 2.2\,K, while all the non-detections are located where the background mean intensity is $\geq$ 2.2\,K (see Table 1).

\section{Conclusions}

We used the SMA to observe a sample of evolved stars in the inner Galactic Bulge in the CO J=2--1 line at 230.5 GHz.  We conclude the following:

\begin{itemize}

\item We detect the high radial velocity star OH 359.943 +0.260 (OH359) in CO J=2--1.  We find it to have a radial velocity of -182.3 ($\pm$ 1.0) km~s$^{-1}$, a shell expansion velocity of V$_{\rm exp}$ = 19.1  ($\pm$ 1.4)  km~s$^{-1}$, and peak emission of 0.36 ($\pm$ 0.02)  Jy.  Using formula A.1 of \citet{ram08}, this implies a gas mass-loss rate of 7.9 ($\pm$ 2.2) $\times$10$^{-5}\,\msunyr$.

\item We detect CO J=2--1 emission from A12 after removing baselines $<$ 25 k$\lambda$.  The best-fit model of its CO J=2--1 emission gives F$_{\nu, \rm max}$ = 0.91 ($\pm$ 0.04) Jy, V$_{rad}$ = 105.0 ($\pm$ 3.0) km~s$^{-1}$, and V$_{\rm exp}$ = 14.3 ($\pm$ 4.2) km~s$^{-1}$.  This implies a gas mass-loss rate of 5.4 ($\pm$ 2.8) $\times$10$^{-5}\,\msunyr$.

\item We also detect CO J=2--1 emission from A51 after removing baselines $<$ 20 k$\lambda$.  Its best-fit CO J=2--1 emission model gives F$_{\nu, \rm max}$ = 3.1 ($\pm$ 0.2) Jy, V$_{rad}$ = -94.8 ($\pm$ 3.0) km~s$^{-1}$, and V$_{\rm exp}$ = 6.0 ($\pm$ 4.2) km~s$^{-1}$, implying a gas mass-loss rate of 3.4 ($\pm$ 3.0) $\times$10$^{-5}\,\msunyr$.  The large uncertainty here is a reflection of the low quality of our detection of the CO J=2--1 emission from A51.

\item We were not successful in detecting the other targets in our sample in the CO J=2--1 line.

\item Modeling of the infrared SEDs of OH359, E51, A12, and A51 obtains dust mass-loss rates of 2.5 ($\pm$ 0.2) $\times$10$^{-7}\,\msunyr$, 8.4 ($\pm$ 1.1) $\times$10$^{-7}\,\msunyr$, 2.4 ($\pm$ 0.1) $\times$10$^{-7}\,\msunyr$, and 2.1 ($\pm$ 0.2) $\times$10$^{-7}\,\msunyr$.  This implies gas-to-dust ratios for OH359, A12, and A51 of 310 ($\pm$ 89), 220 ($\pm$ 110), and 160 ($\pm$ 140), respectively.  For E51, the gas mass-loss rate for E51 of 6$\times$10$^{-5}\,\msunyr$ (A. Winnberg, private communication) and an uncertainty on this gas mass-loss rate that we estimate to be about 2$\times$10$^{-5}\,\msunyr$ (see Section 5.1) implies a gas-to-dust ratio of 71 ($\pm$ 23).  The gas-to-dust ratios for these stars are mostly consistent with their low metallicities, though it is unknown why the gas-to-dust ratio of E51 is lower than that of O-rich AGB stars in the (higher metallicity) solar neighborhood.

\item OH359, A12, and A51 (whose CO J=2--1 emission was detected by us) and E51 \citep[whose CO J=2--1 emission was detected by][]{winn09} are located in regions of lower CO J=2--1 interstellar emission \citep[as measured by][]{saw01}, while our non-detections seemed to occur for stars with larger CO J=2--1 interstellar emission intensities.  This suggests sufficient CO background emission confuses interferometric observations so that faint point sources are obscured and not detected.

\end{itemize}

To attain greater success in observing CO emission from OH/IR and AGB stars in the inner Galactic Bulge, there are a few avenues to explore.  The first is, as \citet{winn09} note, to use a more powerful observing facility like ALMA to observe the CO J=2--1 line.  Alternatively, one may try to observe these stars in a higher line on the CO J $\rightarrow$ J-1 ladder, such as the CO J=3--2 line.  However, perhaps the easiest route is simply to choose targets located in the regions of CO maps such as those of \citet{saw01} or \citet{oka98} with low interstellar CO emission.

\acknowledgements The Submillimeter Array is a joint project 
between the Smithsonian Astrophysical Observatory and the Academia 
Sinica Institute of Astronomy and Astrophysics and is funded by the 
Smithsonian Institution and the Academia Sinica.  The authors would like to acknowledge 
Anders Winnberg for discussions {\it via} email that have been very helpful in interpreting 
our SMA data.  The authors 
wish to thank the anonymous referee for comments that greatly improved 
this manuscript.  This work makes use of 
observations made with the {\it Spitzer Space Telescope}, which is operated by 
the Jet Propulsion Laboratory, California Institute of Technology 
under NASA contract 1407.  This research has made use of the SIMBAD database,
operated at CDS, Strasbourg, France.  This research has made use of the VizieR 
catalogue access tool, CDS, Strasbourg, France.  The original description of the VizieR 
service was published in A\&AS 143, 23.   This research was supported by 
NASA ADP grant NNX11AB06G.

\clearpage

\begin{figure}[t] 
  \epsscale{0.8}
  \includegraphics[scale=0.75,angle=-90]{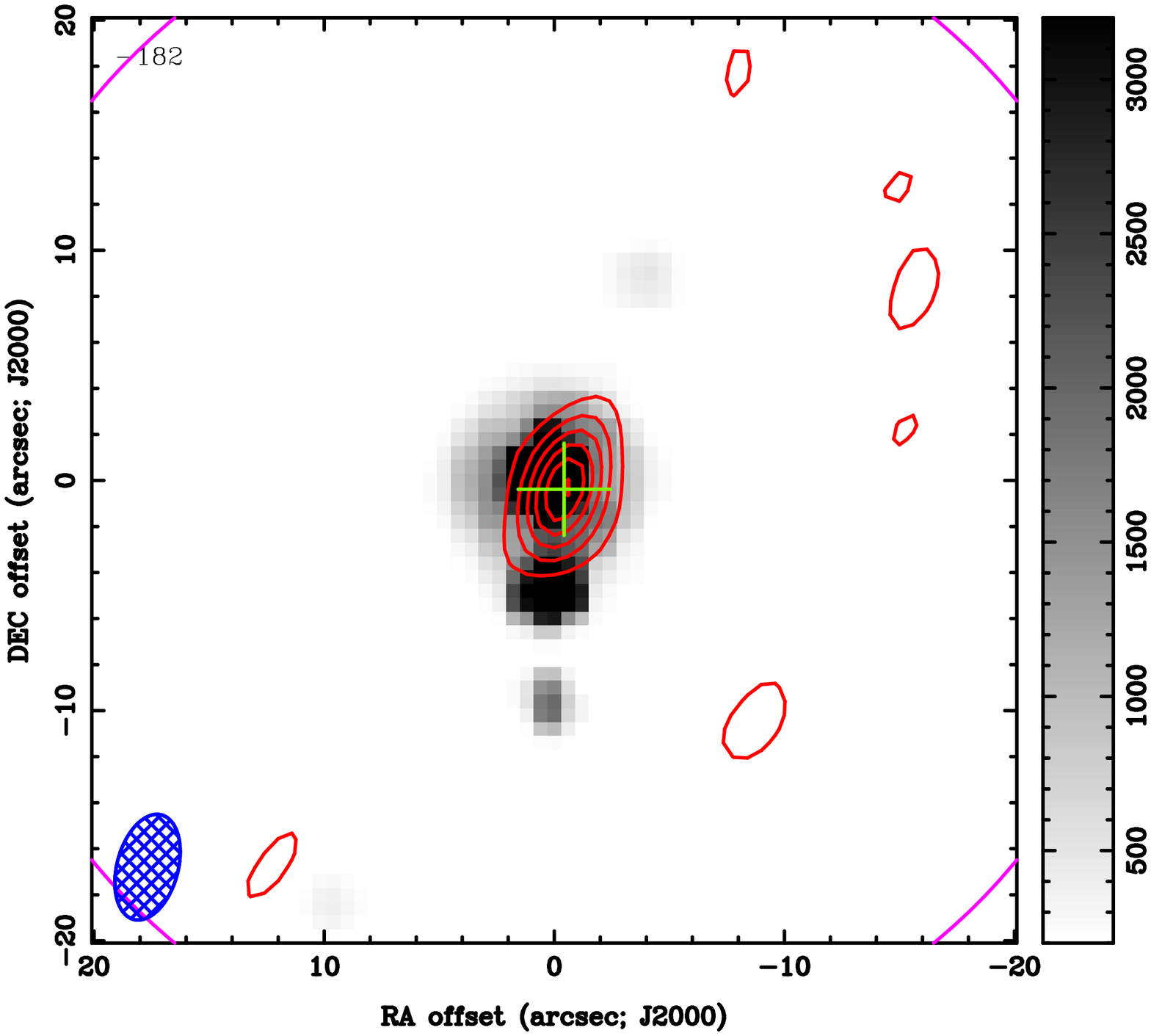}
  \caption[oh359im]{Velocity-binned map for OH 359.943 +0.260, binned over the velocity range of our SMA detection (see Table 2).  The pixel image in the background is a {\it Spitzer}-IRAC 8$\mum$ image from GLIMPSE \citep[][]{ben03,chur09}, AOR \# 13371392, while the contours are the SMA data.  The contour levels are increments of 0.05 Jy/beam, so that the contour level at the peak of the CO emission is 0.3 Jy/beam.  The light green cross indicates the centroid of the source detected in the SMA data.  The size of the image is indicated on the borders of the image.  The greyscale for the {\it Spitzer} pixel image is on the right, and the units for this greyscale are MJy/sr.  The beam size for this data is 4.70$\arcsec$ by 2.66$\arcsec$.  According to the MIRIAD task {\it imstat}, the RMS noise in the square region bounded by (10$\arcsec$,10$\arcsec$), (10$\arcsec$,19$\arcsec$), (19$\arcsec$,10$\arcsec$), and (19$\arcsec$,19$\arcsec$) is 0.02 Jy/beam.  The magenta arcs at the edge of the image are part of a circle 52$\arcsec$ in diameter indicating the SMA primary beam size.}
\end{figure}



\clearpage

\begin{figure}[t] 
  \epsscale{0.8}
  \plotone{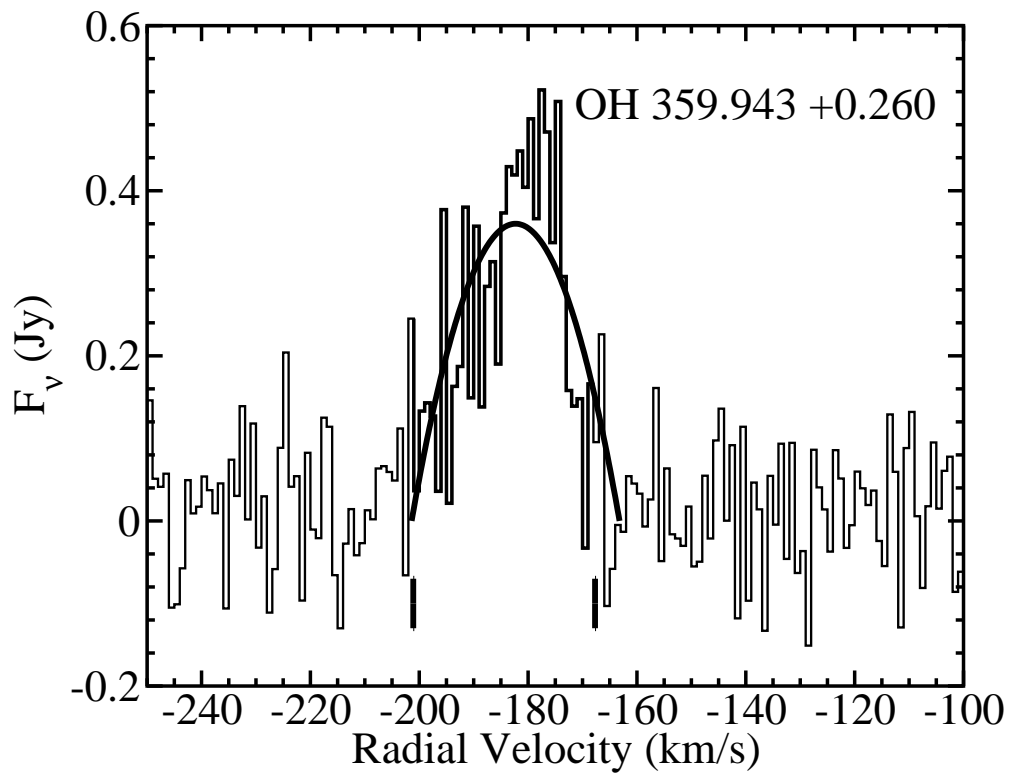}
  \caption[oh359]{The SMA spectrum of the CO J=2-1 line for OH 359.943 +0.260 (histogram), fit by a model (solid line) of CO emission assuming unresolved, optically thick emission from the circumstellar gas shell.  The data within the velocity range over which the model was fit is in bolder lines.  The primary beam correction applied to this spectrum is 1.0003.  The thick short vertical lines indicate the red and blue limits of the OH maser emission for this source detected by \citet{sjou98}.}
\end{figure}

\clearpage

\begin{figure}[t] 
  \includegraphics[scale=0.75,angle=-90]{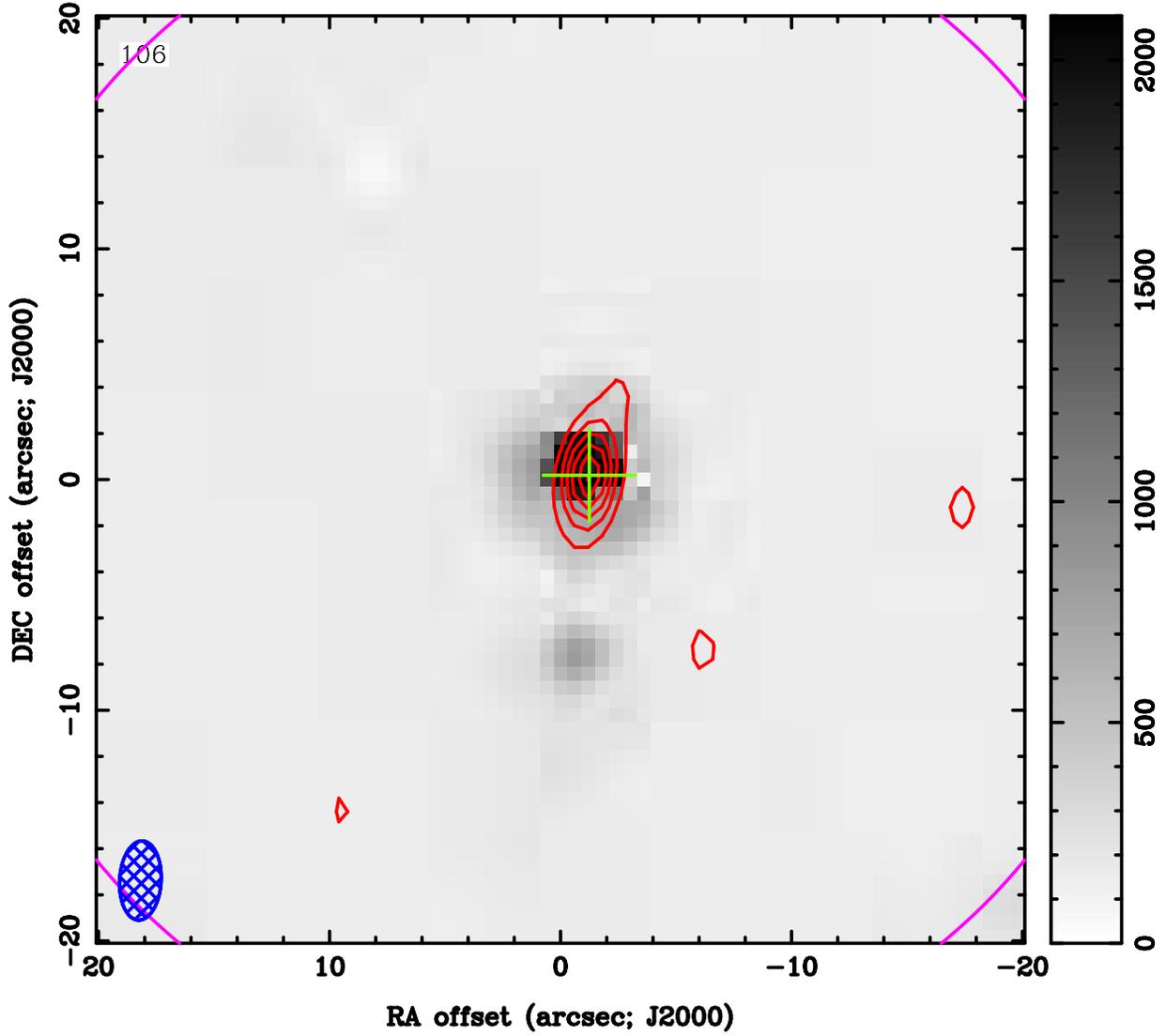}
  \caption[a12im]{Same as Figure 1, but for A12.  The GLIMPSE image is from AOR \# 13370112.  Here the contour 
  levels are increments of 0.1 Jy/beam.  The beam size for this data is 3.43$\arcsec$ by 1.80$\arcsec$.  The RMS noise in the square region bounded by (10$\arcsec$,10$\arcsec$), (10$\arcsec$,19$\arcsec$), (19$\arcsec$,10$\arcsec$), and (19$\arcsec$,19$\arcsec$) is 0.03 Jy/beam.}
\end{figure}



\clearpage

\begin{figure}[t] 
  \epsscale{0.8}
  \plotone{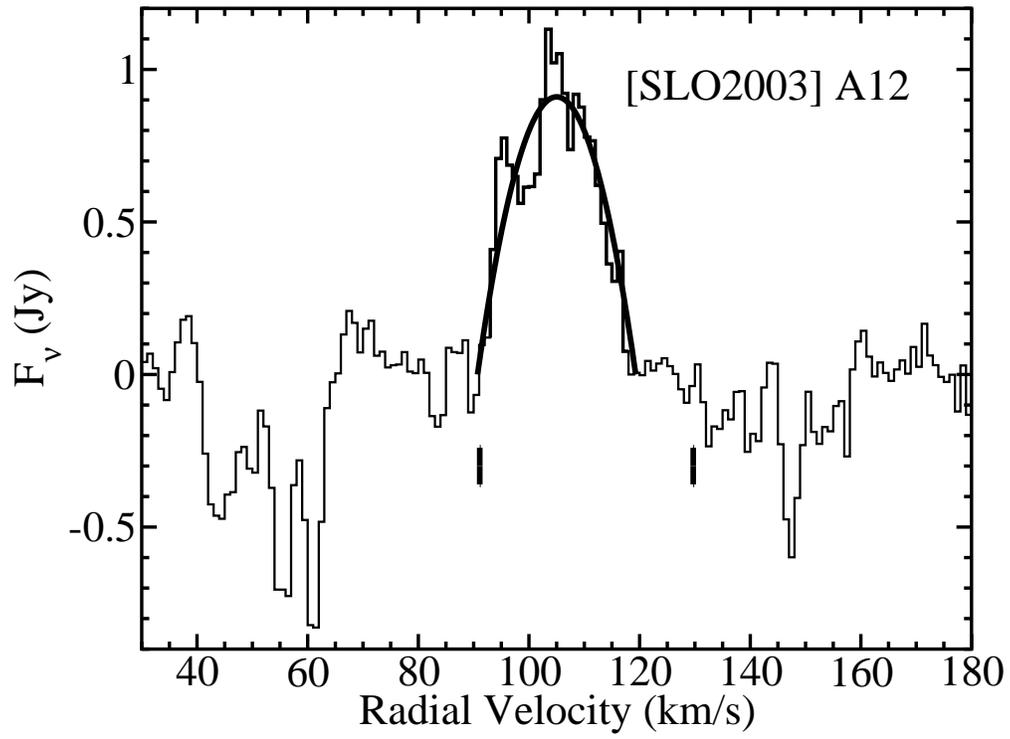}
  \caption[a12]{Same as Figure 2, but for A12.  The primary beam correction applied to this spectrum is 1.00164.}
\end{figure}

\clearpage

\begin{figure}[t] 
  \includegraphics[scale=0.75,angle=-90]{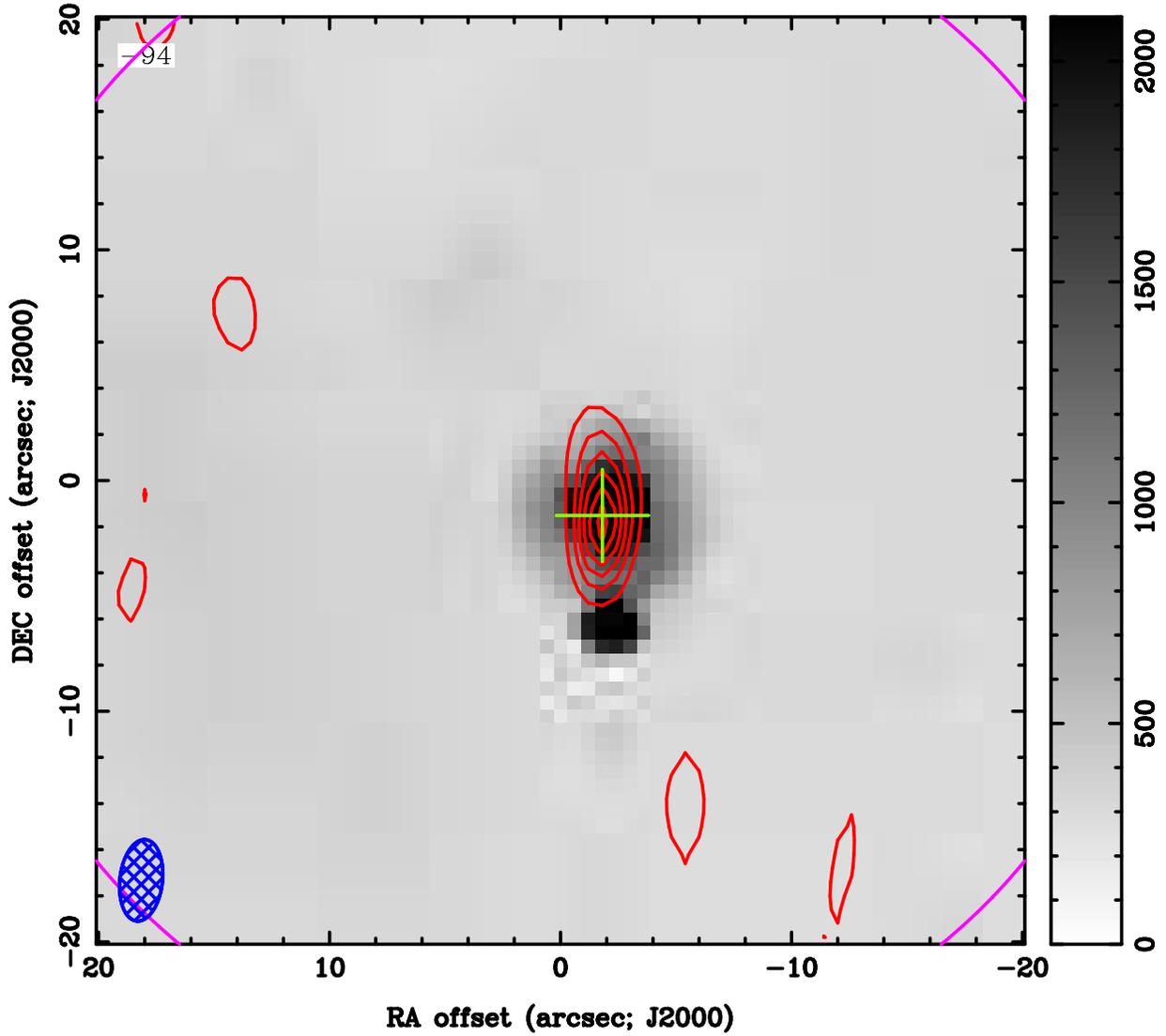}
  \caption[a51im]{Same as Figure 1, but for [SLO2003] A51.  The GLIMPSE image is from AOR \# 13370112.  Here the contour 
  levels are increments of 0.2 Jy/beam.  The beam size for this data is 3.55$\arcsec$ by 1.85$\arcsec$.  The RMS noise in the square region bounded by (10$\arcsec$,10$\arcsec$), (10$\arcsec$,19$\arcsec$), (19$\arcsec$,10$\arcsec$), and (19$\arcsec$,19$\arcsec$) is 0.07 Jy/beam.}
\end{figure}

\clearpage

\begin{figure}[t] 
  \epsscale{0.8}
  \plotone{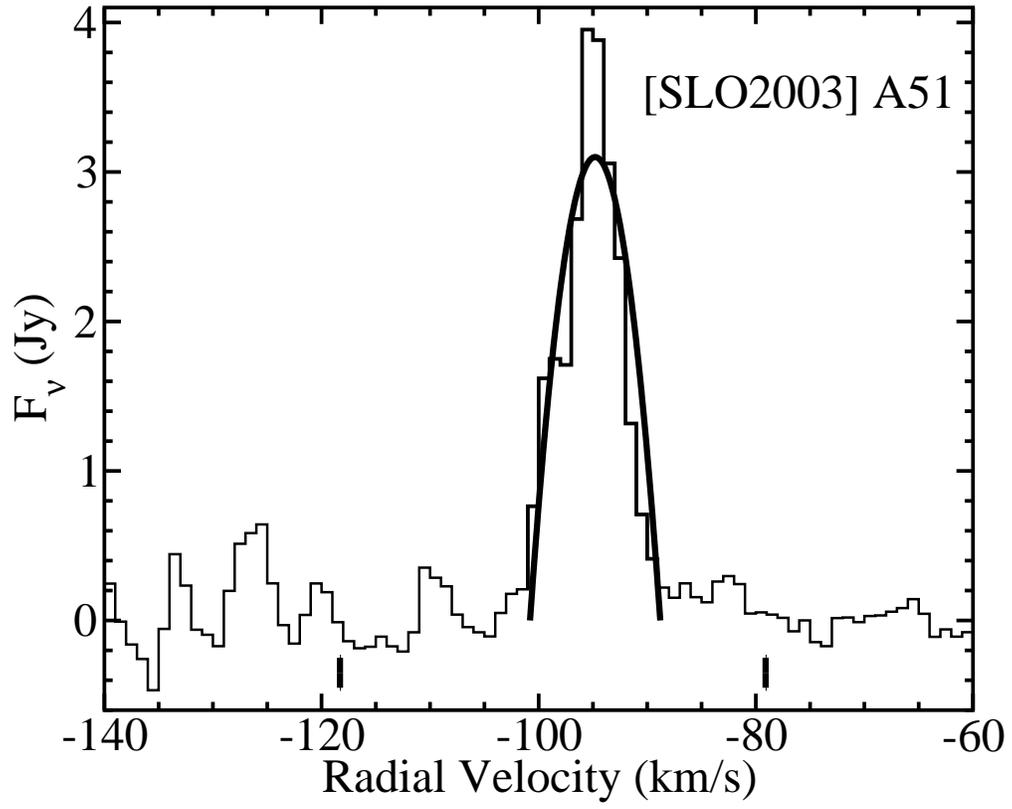}
  \caption[a51]{Same as Figure 2, but for [SLO2003] A51.  The primary beam correction applied to this spectrum is 1.00576.}
\end{figure}

\clearpage

\begin{figure}[t] 
  \includegraphics[scale=0.75,angle=-90]{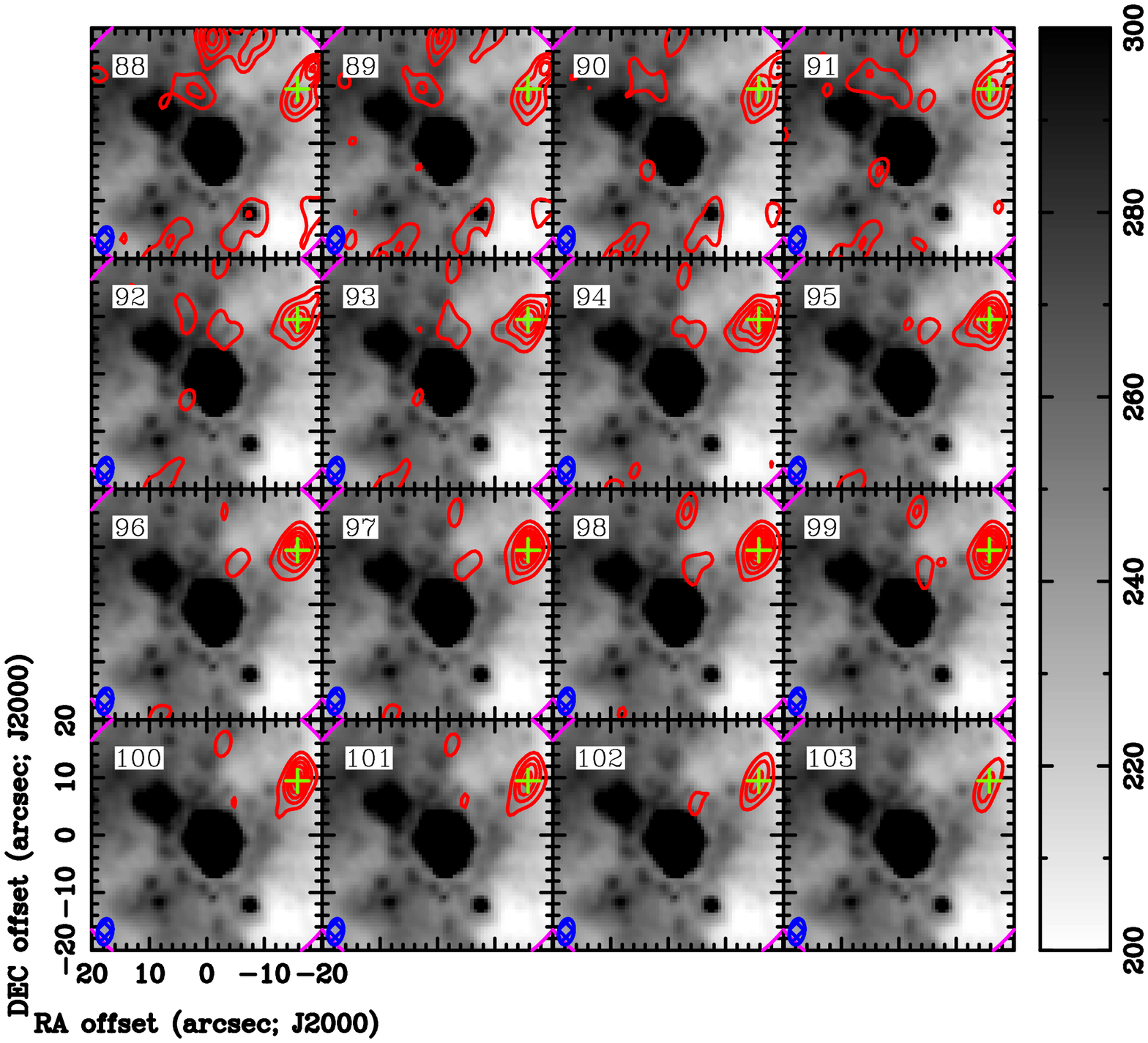}
  \caption[a10im]{Same as Figure 1, but for an off-center source in the [SLO2003] A10 field.  The GLIMPSE image is from AOR \# 13370112.  Here the contour 
  levels are increments of 0.2 Jy/beam.  The beam size for this data is 4.55$\arcsec$ by 2.62$\arcsec$.  The RMS noise at the 96 km~s$^{-1}$ channel, 0.04 Jy/beam, is representative of that of the nearby velocity channels.  This figure is available online only.}
\end{figure}

\clearpage

\begin{figure}[t] 
  \epsscale{0.8}
  \plotone{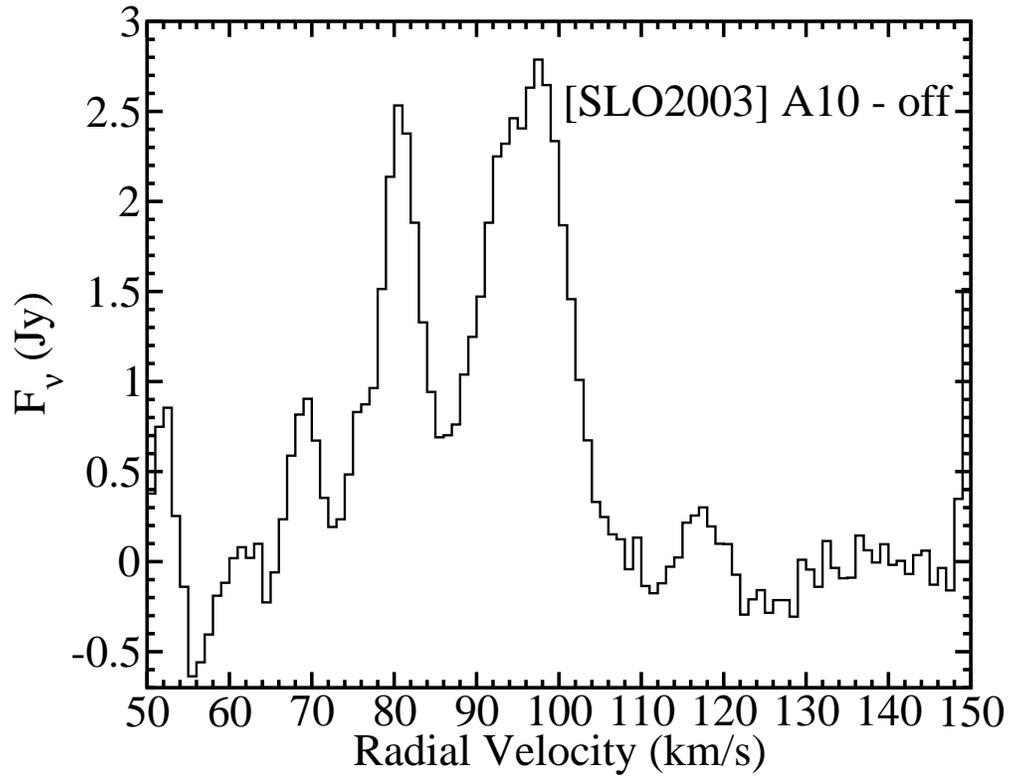}
  \caption[a10]{Same as Figure 2, but for the off-center source in the [SLO2003] A10 field.  The primary beam correction applied to this spectrum is 1.41496.    This figure is available online only.}
\end{figure}

\clearpage

\begin{figure}[t] 
  \includegraphics[scale=0.75,angle=-90]{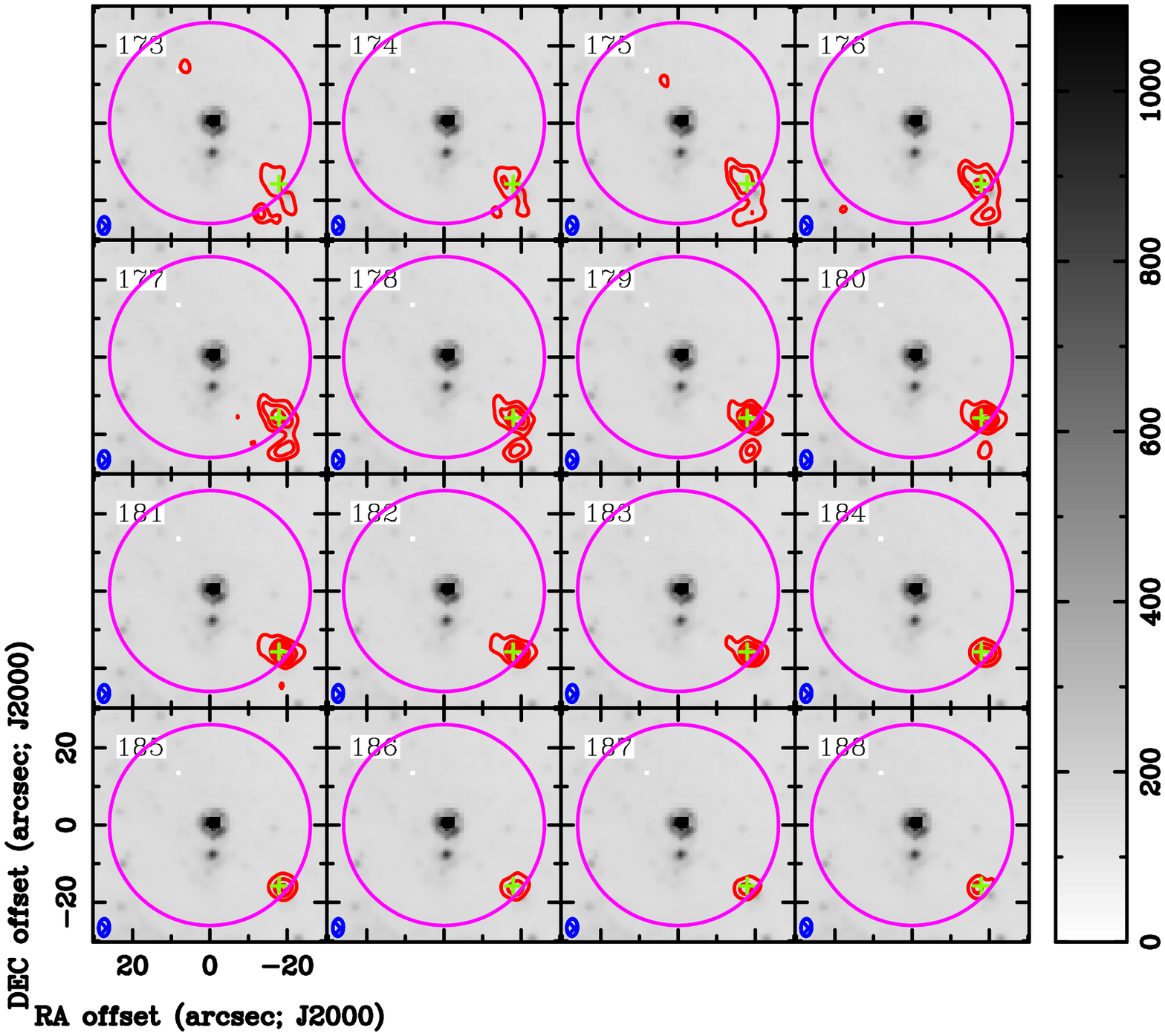}
  \caption[a12ntim]{Same as Figure 1, but for an off-center source in the [SLO2003] A12 field, and the images here are 60x60$\arcsec$.  The GLIMPSE image is from AOR \# 13370112.  Here the contour levels are increments of 0.3 Jy/beam.  The beam size for this data is 4.49$\arcsec$ by 2.74$\arcsec$.  The RMS noise at the 181 km~s$^{-1}$ channel, 0.05 Jy/beam, is representative of that of the nearby velocity channels.  This figure is available online only.}
\end{figure}

\clearpage

\begin{figure}[t] 
  \epsscale{0.8}
  \plotone{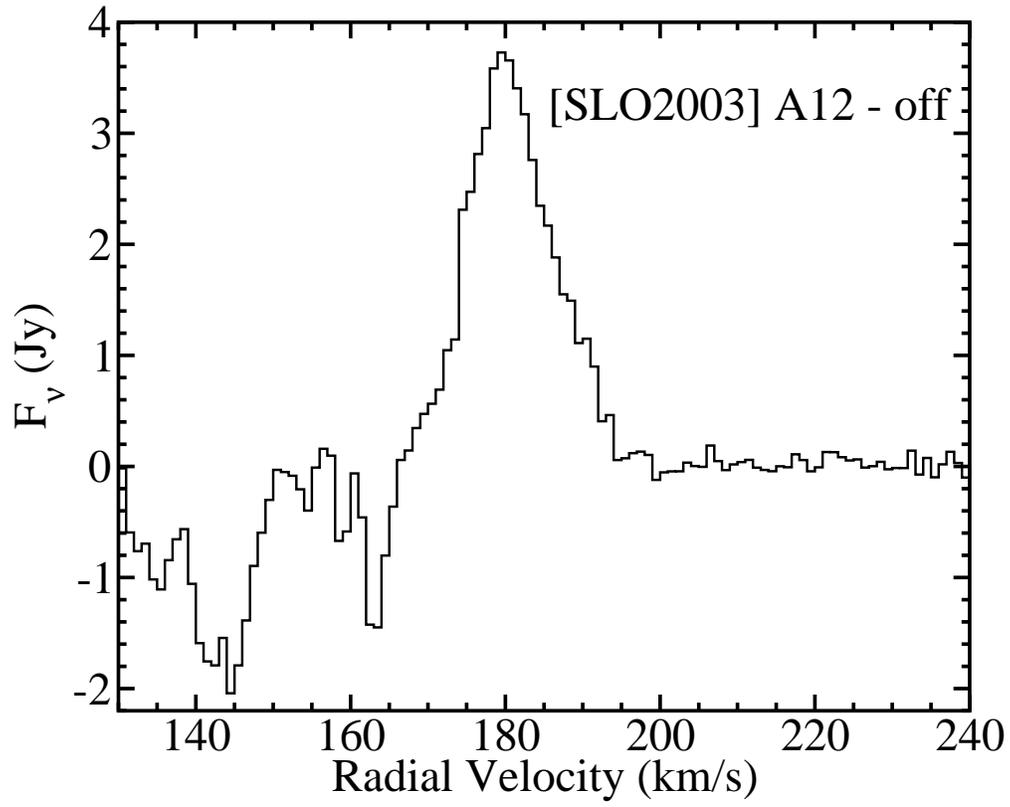}
  \caption[a12nt]{Same as Figure 2, but for the off-center source in the [SLO2003] A12 field.  The primary beam correction applied to this spectrum is 1.7919.  This figure is available online only.}
\end{figure}

\clearpage

\begin{figure}[t] 
  \includegraphics[scale=0.75,angle=-90]{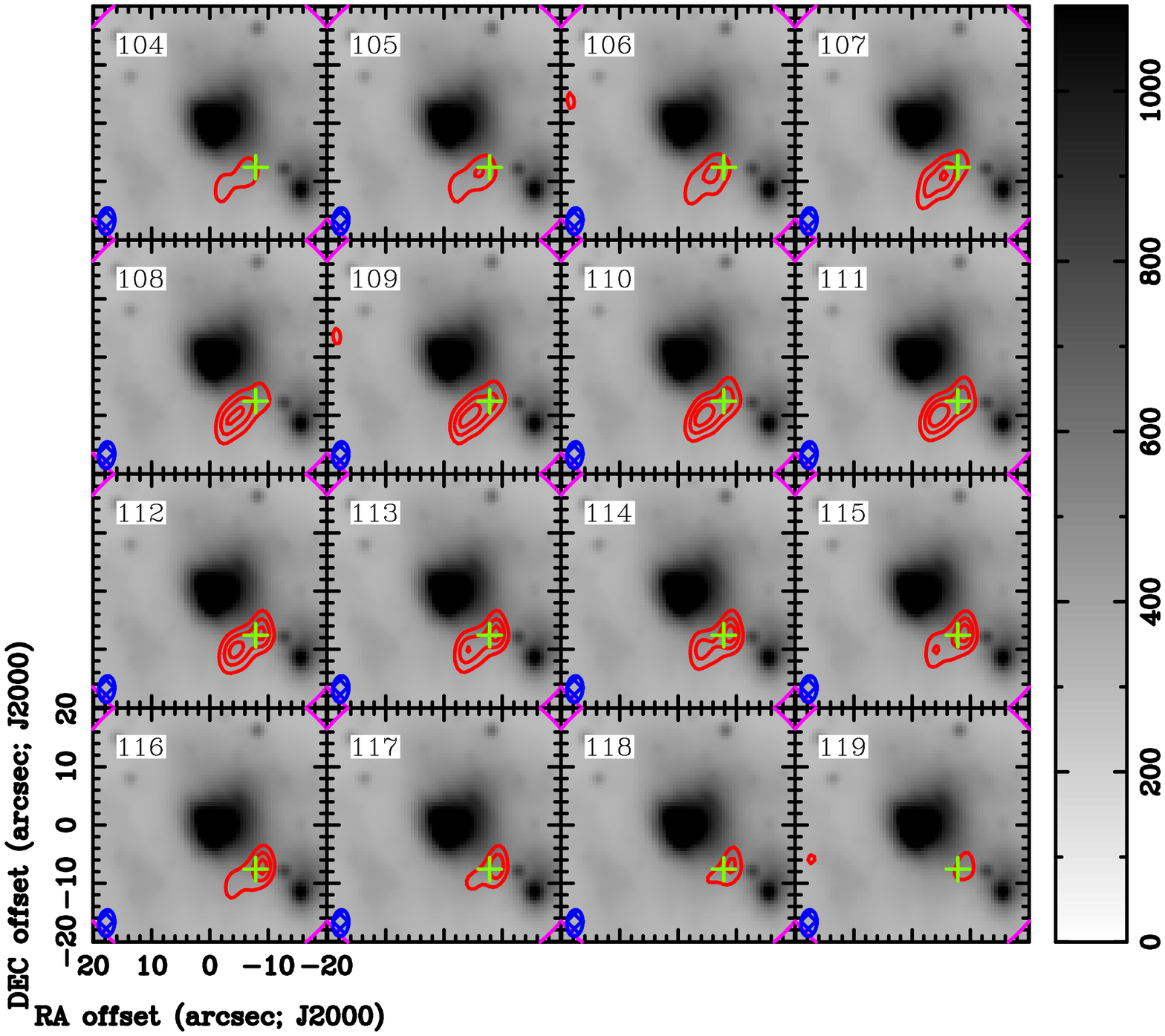}
  \caption[a27im]{Same as Figure 1, but for an off-center source in the [SLO2003] A27 field.  The GLIMPSE image is from AOR \# 13368832.  Here the contour 
  levels are increments of 0.5 Jy/beam.  The beam size for this data is 4.53$\arcsec$ by 2.71$\arcsec$.  The RMS noise at the 112 km~s$^{-1}$ channel, 0.06 Jy/beam, is representative of that of the nearby velocity channels.  This figure is available online only.}
\end{figure}

\clearpage

\begin{figure}[t] 
  \epsscale{0.8}
  \plotone{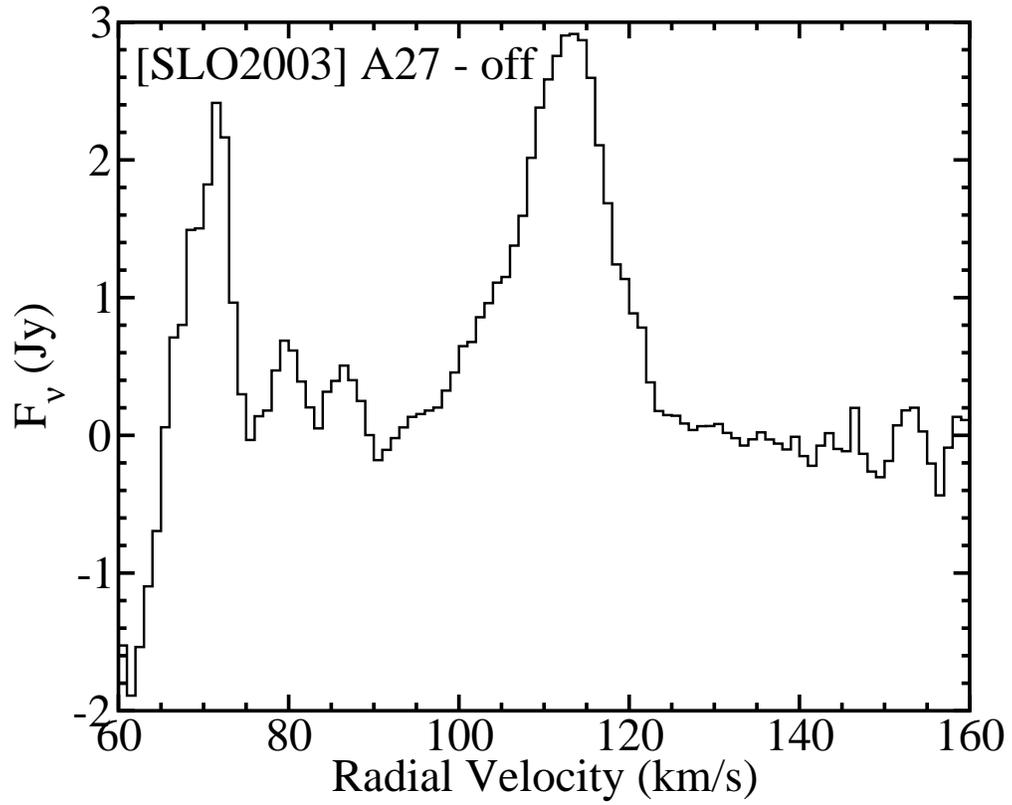}
  \caption[a27]{Same as Figure 2, but for the off-center source in the [SLO2003] A27 field.  The primary beam correction applied to this spectrum is 1.1387.  This figure is available online only.}
\end{figure}

\clearpage

\begin{figure}[t] 
  \includegraphics[scale=0.5]{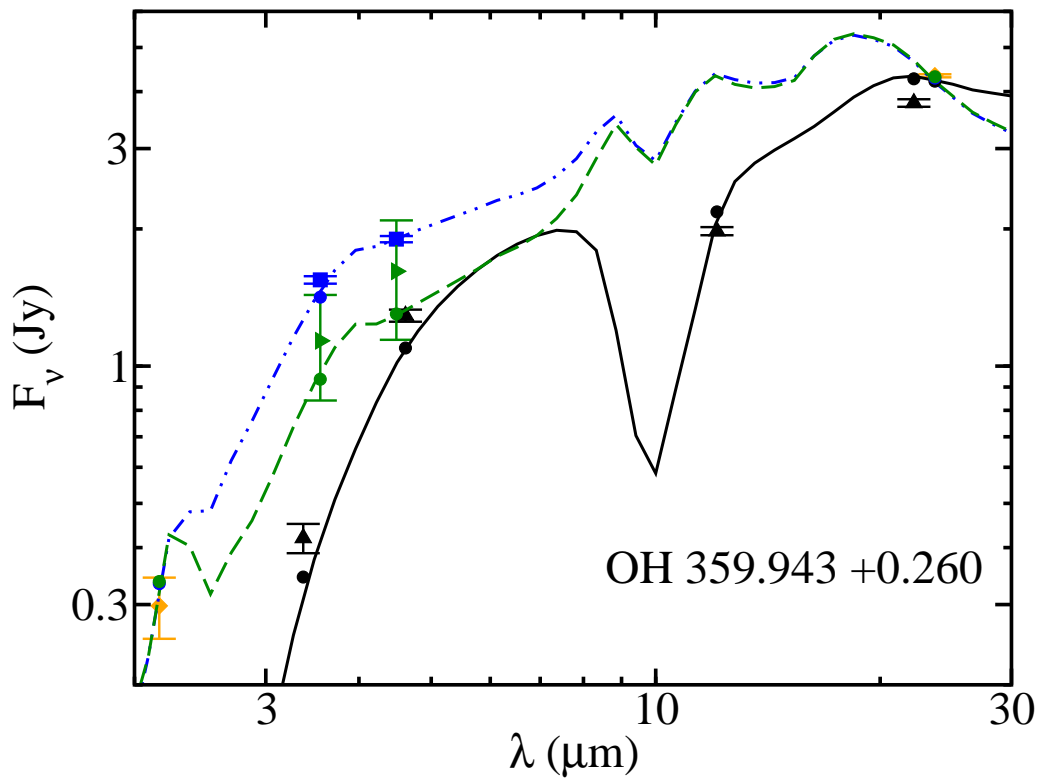}
  \caption[oh359sed]{The SED for OH 359.943 +0.260.  
The black triangles with error bars are observed WISE data, the blue squares with error bars are the observed \citet{rmz08} 
photometry, the green triangles pointing right with error bars are the observed GLIMPSE photometry, and the orange diamonds 
with error bars are the photometry common to all 3 epochs.  The black solid line and black circles are the best-fit model spectrum 
and synthesized photometry, respectively, for the data set with WISE photometry; the blue dash-dot-dot line and blue circles are the 
best-fit model spectrum and synthesized photometry, respectively, for the data set with \citet{rmz08} photometry; and the green 
dashed line and green circles are the best-fit model spectrum and synthesized photometry, respectively for the data set with 
GLIMPSE photometry.}
\end{figure}

\clearpage

\begin{figure}[t] 
  \includegraphics[scale=0.5]{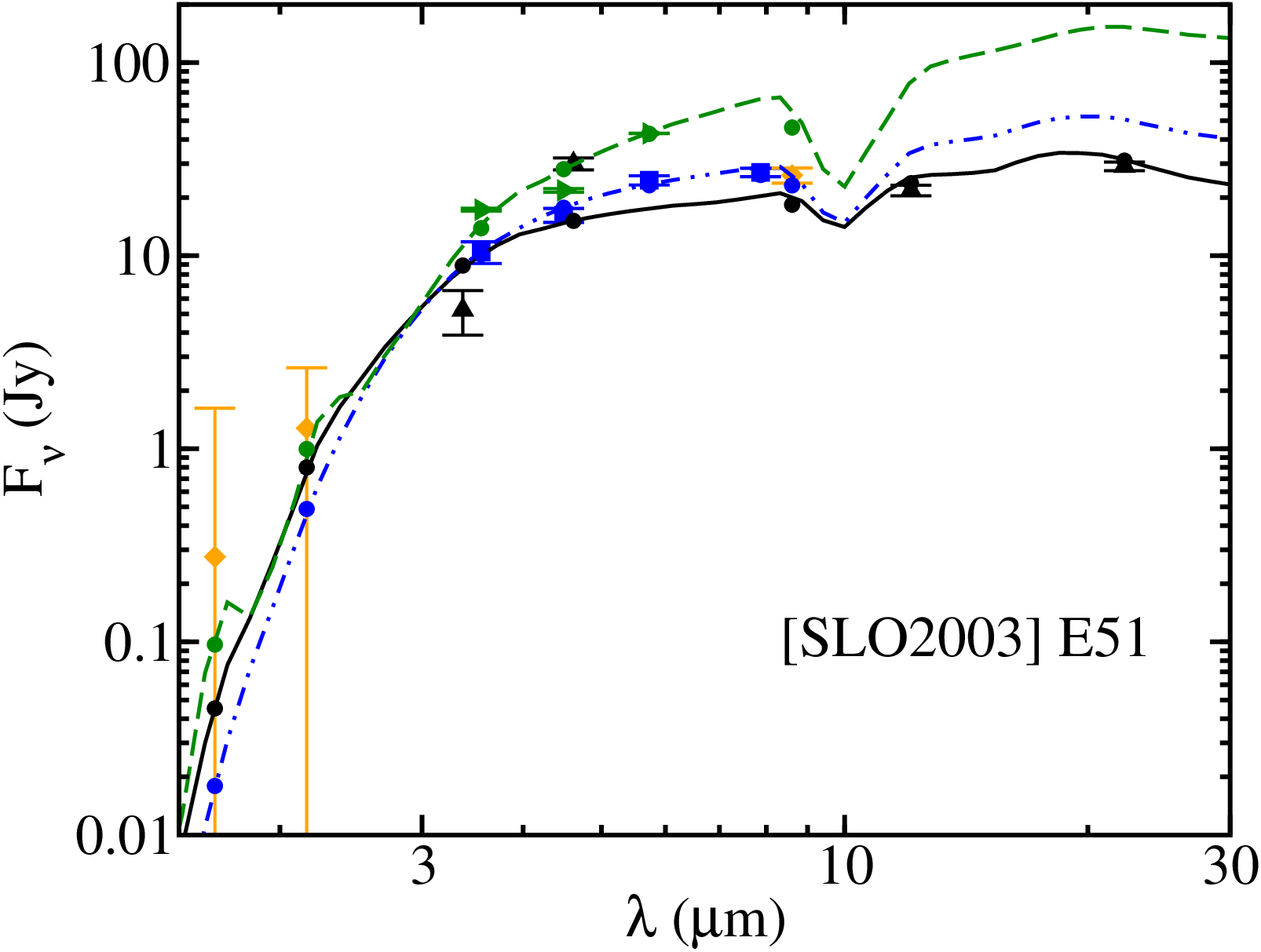}
  \caption[e51sed]{Same as Figure 13, but for E51.}
\end{figure}

\clearpage

\begin{figure}[t] 
  \includegraphics[scale=0.5]{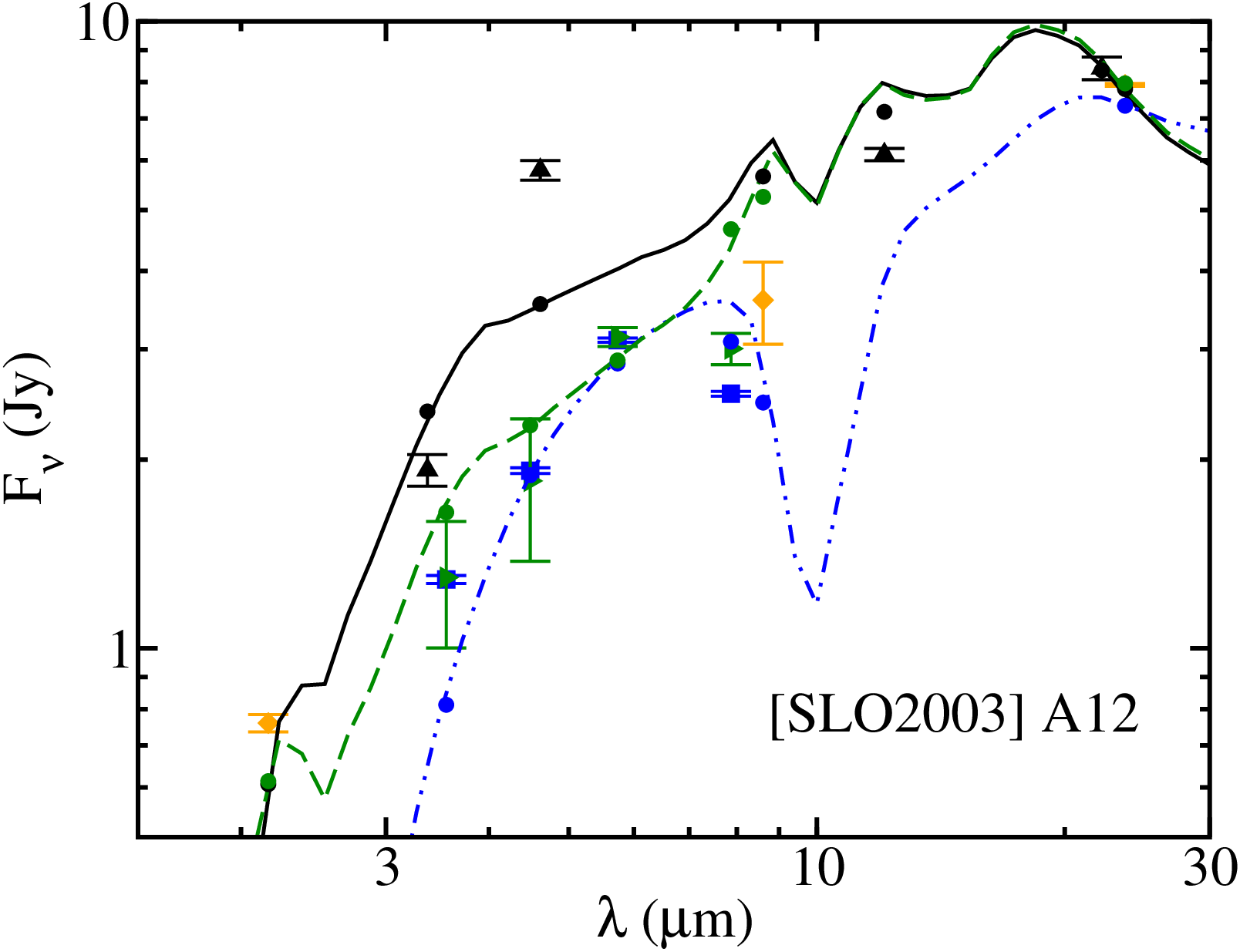}
  \caption[a12sed]{Same as Figure 13, but for A12.}
\end{figure}

\clearpage

\begin{figure}[t] 
  \includegraphics[scale=0.5]{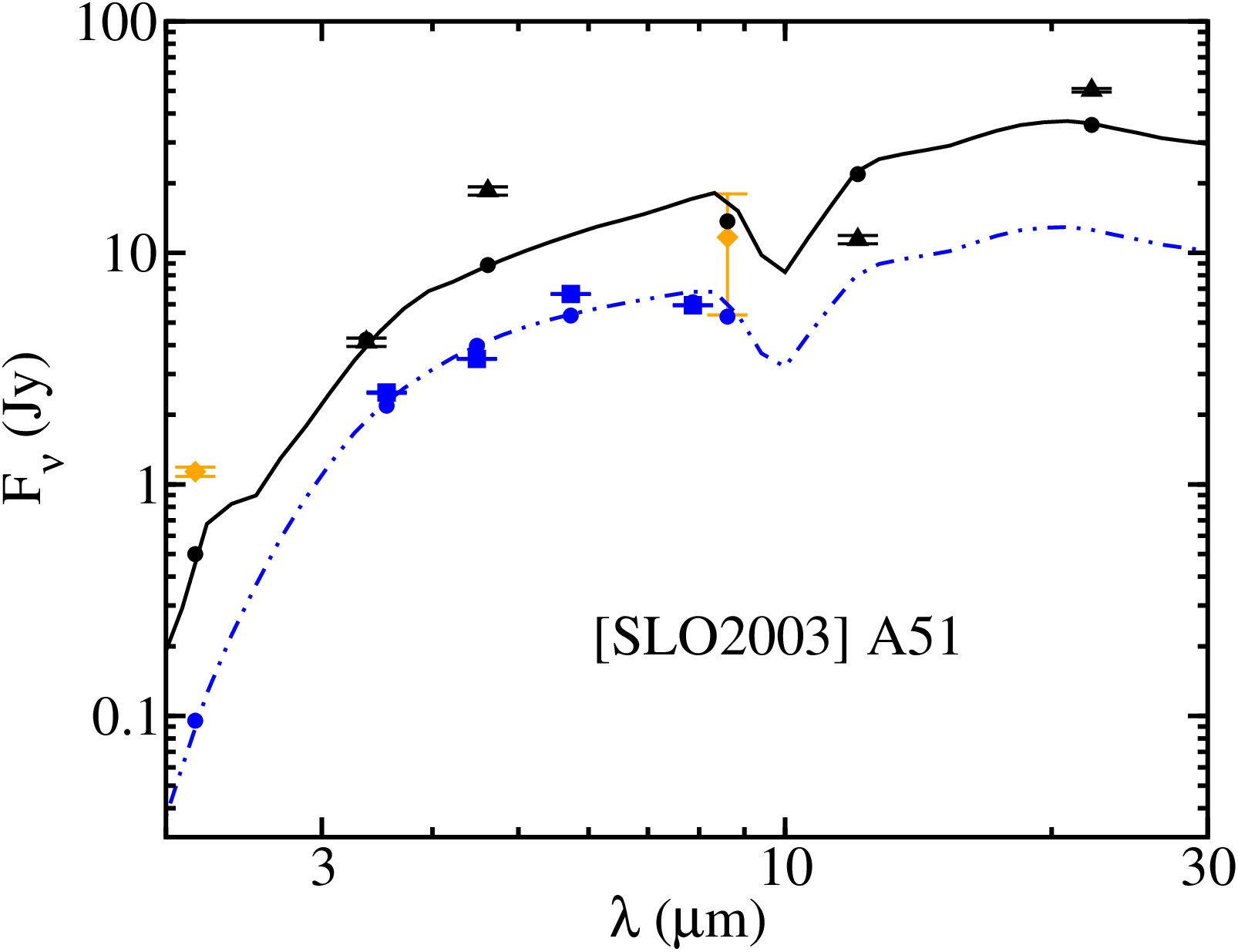}
  \caption[a51sed]{Same as Figure 13, but for A51.}
\end{figure}

\clearpage

\begin{landscape}
\begin{table}[h,t]
{
\caption[sample]{Information from Literature \label{table1}}
\begin{tabular}{lccccccccc}
\hline \hline
          Name
          & R. A.
          & Dec
          & V$_{\rm blue}$
          & V$_{\rm radial}$
          & V$_{\rm red}$
          &
          &
          & 
          & CO J=2--1\\
          
          & (J2000)
          & (J2000)
          & (km~s$^{-1}$)
          & (km~s$^{-1}$)
          & (km~s$^{-1}$)
          & Type
          & Alias
          & Ref.$^{a}$
          & B.G. (K)$^{b}$\\
\hline
OH359 & 17h44m28.2s & -28d50m55.8s & -201.1 & -184.4 & -167.7 & OH/IR & OH 359.943 +0.260 & 1 & 0.1\\
A7 & 17h44m34.4s & -29d10m38.5s & -42.8 & -24.6 & -6.4 & OH/IR & OH 359.676 +0.069 & 1 & 2.7\\
A10 & 17h44m39.7s & -29d16m45.9s & -92.1 & -73.3 & -54.4 & OH/IR & OH 359.598 +0.000 & 1 & 4.0\\
A12 & 17h44m44.5s & -29d05m38.3s & +91.1 & +110.4 & +129.7 & OH/IR & OH 359.765 +0.082 & 1 & 0.8\\
A27 & 17h44m57.0s & -29d05m57.3s & \nodata & \nodata & \nodata & AGB & ISOGAL-P J174457.0-290557 & \nodata & \nodata\\
A29 & 17h44m57.8s & -29d20m42.5s & -74.1 & -55.5 & -36.9 & OH/IR & OH 359.58 -0.09 & 2 & 3.2\\
A51 & 17h45m14.0s & -29d15m27.4s & -118.3 & -98.7 & -79.1 & OH/IR & OH 359.681 -0.095 & 1 & 2.2\\
A52 & 17h45m14.3s & -29d07m20.8s & -90.5 & -71.2 & -51.9 & OH/IR & OH 359.797 -0.025 & 1 & 2.2\\
\hline
\end{tabular}
\tablenotetext{a}{\footnotesize References for the OH detections of our sample.  The references are: (1) \citet{sjou98}; (2) \citet{lqv92}.}
\tablenotetext{b}{\footnotesize The background mean intensity at the location of the star, in Kelvins, as measured by \citet{saw01}, interpolated 
at the location of our target in the \citet{saw01} map.  The source observed by \citet{winn09}, E51, has a background CO J=2--1 mean intensity of 1.8\,K.}
}
\end{table}
\end{landscape}

\clearpage

\begin{landscape}
\begin{table}[h,t]
{
\caption[sample]{Observations and Analysis \label{table2}}
\begin{tabular}{lcccccccc}
\hline \hline
          Target
          & Date
          & Observed
          & Observed
          & V$_{\rm rad}$
          & V$_{\rm exp}$
          & F$_{\nu, \rm max}$
          & I$_{\rm co}$
          & $\mdot$\\
          
          & Observed
          & R.A. (J2000)
          & Dec (J2000)
          & (km~s$^{-1}$)
          & (km~s$^{-1}$)
          & (Jy)
          & (K$\times$km~s$^{-1}$)
          & (10$^{-5}\,\msunyr$)\\
\hline
OH359 & 15 May 2012 & 17h44m28.2s & -28d50m56.2s & -182.3 ($\pm$ 1.0) & 19.1 ($\pm$ 1.4) & 0.36 ($\pm$ 0.02) & 98.1 ($\pm$ 33.0) & 7.9 ($\pm$ 2.2)\\
A12 & 3 May 2010 & 17h44m44.4s & -29d05m38.1s & 105.0 ($\pm$ 3.0) & 14.3 ($\pm$ 4.2) & 0.91 ($\pm$ 0.04) & 145.5 ($\pm$ 88.1) & 5.4 ($\pm$ 2.8)\\
A51 & 11 May 2010 & 17h45m13.9s & -29d15m28.9s & -94.8 ($\pm$ 3.0) & 6.0 ($\pm$ 4.2) & 3.1 ($\pm$ 0.2) & 126.0  ($\pm$ 127.0) & 3.4 ($\pm$ 3.0)\\
\hline
\end{tabular}
\tablecomments{\footnotesize The uncertainties on the various quantities listed in this table are determined as we describe in Section 4.}
}
\end{table}
\end{landscape}

\begin{table}[h,t]
{
\caption[sample]{Dust Modeling  \label{table3}}
\begin{tabular}{lcccc}
\hline \hline
          
          & OH359
          & E51
          & A12
          & A51\\
\hline
[Fe/H]$^{a}$ & -0.81 & -0.92 & -0.82 & -0.38\\
2MASS src\#$^{b}$ & \nodata & 17443496 & 17444443 & 17451393\\
 & \nodata & -2904356 & -2905379 & -2915279\\
DENIS id\#$^{c}$ & 98816 & \nodata & \nodata & \nodata\\
R08 src\#$^{d}$ & 0334881 & 0352203 & 0376998 & 0454462\\
GLIMPSE src\#$^{e}$ & G359.9429 & G359.7619 & G359.7650 & \nodata\\
 & +00.2603 & +00.1202 & +00.0817 & \nodata\\
MIPS rec\#$^{f}$ & 105951 & \nodata & 102210 & \nodata\\
AKARI \#$^{g}$ & \nodata & 200500200 & 200500470 & 200501468\\
WISE$^{h}$ & J174428.18 & J174434.95 & J174444.44 & J174513.877\\
 & -285056.1 & -290435.5 & -290537.9 & -291528.6\\
$<$mag$>^{i}$ & 9.87 & 4.5 & 10.4 & 11.25\\
Amp$^{i}$ & 2.21 & 1.82 & 2.22 & 2.95\\
Band$^{i}$ & $K$ & $L'$ & $K$ & $K$\\
$\mdot_{\rm dust}$ ($\times$10$^{-7}\,\msunyr$) & 2.5 ($\pm$ 0.2) & 8.4 ($\pm$ 1.1) & 2.4 ($\pm$ 0.1) & 2.1 ($\pm$ 0.2)\\
M$_{\rm gas}$/M$_{\rm dust}$ & 310 ($\pm$ 89) & 71 ($\pm$ 23)$^{j}$ & 220 ($\pm$ 110) & 160 ($\pm$ 140)\\
\hline
\end{tabular}
\tablecomments{\footnotesize The id, reference, and record numbers we list here for 2MASS, DENIS, {\it Spitzer} IRAC 
\citep[both GLIMPSE and][]{rmz08}, {\it Spitzer} MIPS, AKARI, and WISE database entries are those found in Vizier in SIMBAD.}
\tablenotetext{a}{\footnotesize The metallicities listed here are determined using Solution 1 of \citet{ram00} and NIR line equivalent widths from \citet{schul03} for E51, A12, and A51, and \citet{vanhol06} for OH359.}
\tablenotetext{b}{\footnotesize See \citet{skrut06}.  The source designation numbers are for the entries in II/246/out in Vizier.}
\tablenotetext{c}{\footnotesize See \citet{epcht99}.  The record number is for the entry in II/243/psc in Vizier.}
\tablenotetext{d}{\footnotesize See \citet{rmz08}.  The source numbers are for the entries in II/295/SSTGC, from \citet{rmz08}, in Vizier.}
\tablenotetext{e}{\footnotesize See \citet{ben03} and \citet{chur09}.  The record numbers are for the GLIMPSE entries in Vizier.}
\tablenotetext{f}{\footnotesize See \citet{hinz09}.  The record numbers are for the entries in J/ApJS/181/227/table2 in Vizier.}
\tablenotetext{g}{\footnotesize See \citet{ishi10}.  The source id numbers are for the entries in II/297/irc in Vizier.}
\tablenotetext{h}{\footnotesize See \citet{wri10}.  The WISE All-Sky Release Catalog names are for the entries in II/311/wise in Vizier.}
\tablenotetext{i}{\footnotesize ``$<$mag$>$'', ``Amp'', and ``band'' are the average magnitude, amplitude, and photometric band for which the variability of the star was measured.  For E51, the variability information comes from \citet{jones94} for target OH 359.76 +0.12, while for OH359, A12, and A51, the information comes from \citet{wood98}, for targets 64-28, 35-18, and 23-12, respectively.}
\tablenotetext{j}{\footnotesize As we note in the text, for the gas-to-dust ratio of E51, we use the gas mass-loss rate of 6$\times$10$^{-5}\,\msunyr$ (A. Winnberg, private communication).  Also, we estimate the uncertainty on the E51 gas mass-loss rate by propagating the uncertainties given by \citet{winn09} for E51's CO J=2--1 F$_{\nu, \rm max}$, V$_{rad}$, and V$_{\rm exp}$.  This gas mass-loss rate uncertainty and that of its dust mass-loss rate (from our infrared SED-fitting) are propagated to obtain the gas-to-dust ratio uncertainty for E51.}
}
\end{table}

\end{document}